\documentclass[apj,twocolumn,twocolappendix,numberedappendix,appendixfloat]{openjournal}

\usepackage{graphicx}
\usepackage{amsmath}
\usepackage{amssymb}

\usepackage{siunitx}
\usepackage{comment}
\usepackage{multirow}
\usepackage[breaklinks,colorlinks,citecolor=blue,urlcolor=blue,linkcolor=blue]{hyperref}
\usepackage{xcolor}
\usepackage{ragged2e}

\def\micron{$\mu$m}
\def\bh1{N6946-BH1}

\begin{document}

\title{The neighboring stars of N6946-BH1 and the observational characteristics of failed supernovae}

\author{\vspace{-1.3cm}R.~For\'es-Toribio\,$^{1,2}$}
\email[Email: ]{forestoribio.1@osu.edu}
\author{C.~S.~Kochanek\,$^{1,2}$}

\affiliation{$^{1}$Department of Astronomy, The Ohio State University, 140 West 18th Avenue, Columbus, OH 43210, USA}
\affiliation{$^{2}$Center for Cosmology and Astroparticle Physics, The Ohio State University, 191 W. Woodruff Avenue, Columbus, OH 43210, USA}

\begin{abstract}
Stellar collapse models predict that some stars more massive than $\sim$15$M_\odot$ may collapse directly to a black hole, sometimes with a weak optical transient, a phenomenon known as a failed supernova. Detecting such events is challenging, but searches of vanishing stars have found two promising candidates, N6946-BH1 and M31-2014-DS1. We re-analyze the JWST NIRCam and MIRI images of N6946-BH1 to characterize the remnant emission of the object and its surrounding sources. We found four near-infrared stellar neighbors within approximately 0\farcs1 that are not related to the mid-infrared emission of the candidate. The SED of N6946-BH1 is well modeled by a $\log(L/L_\odot)$=$4.727_{-0.025}^{+0.026}$ (at 95\% confidence) source obscured by a silicate dust shell with a maximum grain size of $\sim$3~$\mu$m and producing negligible emission at $\lesssim$2 \micron. Given proposals that failed SN candidates are actually stellar mergers, we model the progenitor and remnant emission of four Galactic and seven extragalactic stellar mergers and compare their properties with those of the failed supernova candidates. We found that the merger remnants are 10-100 times more luminous than their progenitors at these late phases while the remnants of failed supernovae are $\sim$10 times dimmer than their progenitors. Axisymmetric, disky dust distributions cannot explain the factor of $\sim$100 or more difference in the ratios of the progenitor and remnant luminosities.
\end{abstract}
\keywords{Core-collapse supernovae (304), Massive stars (732), Black holes (162), Stellar mergers (2157)}

\section{Introduction}\label{sec:intro}

Understanding the deaths of massive stars is important for galactic chemical evolution and the regulation of star formation. They are also the source of black holes (BH) and neutron stars (NS) while also determining the survival of binaries which may later interact or merge as gravitational wave (GW) sources. Both observations and stellar evolution theories agree that stars more massive than $\sim$8$M_\odot$ undergo core collapse \citep[see, e.g.,][]{Mezzacappa2005,Botticella2012,VanDyk2012,Maund2014,Burrows2021}. The lower part of this mass range ($\lesssim$15$M_\odot$) dominates the rates and the neutrino mechanism likely leads to successful supernova (SN) explosions and the formation of a NS for these stars.

At higher masses the situation is more complicated. Modern surveys of the ``explodability'' of massive stars find a complex landscape of explosions producing NS and failed explosions producing BH with very few examples of ``fall back'' formation of a BH in a successful SN \citep[e.g.,][]{OConnor2011,Ugliano2012,Pejcha2015,Ertl2016,Sukhbold2016,Luo2025,Ugolini2025}. These models suggest that 10\%-30\% of core collapses lead to a failed SN.

Observational searches for SN progenitors have led to arguments both in favor and against these theories. For example, there are arguments for an absence of Type II SNe with progenitor initial masses $\gtrsim$18$M_\odot$, whereas the population of red supergiant (RSG) stars should extend to higher masses, up to $\sim$25$M_\odot$ \citep[see][and references therein]{Smartt2015}. This is called the ``red supergiant problem'' although there are arguments against the existence of a problem \citep[e.g.,][]{Davies2020-1,Davies2020-2} and counterarguments \citep[e.g.,][]{Kochanek2020}. Inferring supernova progenitor masses is difficult due to the generally limited archival wavelength coverage, although this should change in the near future with the newly launched Nancy Grace Roman Space Telescope\footnote{http://roman.gsfc.nasa.gov/} \citep{Spergel2015,Akeson2019}.

Another argument in favor of failed SN is the mass distribution of black holes. The merging BH binaries found in LIGO-Virgo-KAGRA range from $\sim$6 to 137 $M_\odot$ \citep{LVK2025}. The lower end of the distribution is noticeably higher than the maximum neutron star mass \citep[$M\lesssim3M_\odot$,][]{Kalogera1996}, suggesting a gap between NS and BH remnant masses. Although GW detections are biased towards higher BH masses, this mass gap is also observed for Galactic NS and BH binaries \citep[e.g.,][]{Ozel2012,Kreidberg2012}. The remnant mass gap cannot be reproduced if BHs are only formed by ``fall back'', as this mechanism produces a continuous remnant mass distribution. On the other hand, the failed SN mechanism will form BHs with the masses of their progenitor's helium core, $\sim$5-10$M_\odot$, naturally producing this gap \citep{Kochanek2014SNmass}.

The definitive proof of failed SN would be to identify these events directly rather than trying to infer their occurrence from the properties of SN and compact objects. Failed SNe are predicted to have gravitational wave \citep{Vartanyan2023,Powell2025} and neutrino \citep{Liebendorfer2004,Kuroda2023} emissions which will unambiguously identify them as such. However, current and near-future GW and neutrino detectors are limited to Galactic events, leading to expected rates of one per several centuries. \cite{Kochanek2008} pointed out an alternative method to find vanishing stars possibly with a weak optical transient that can be used for nearby galaxies ($\lesssim$10~Mpc) with current 8-m telescopes, allowing the detections of examples over decades rather than centuries.

This has led to a range of theoretical studies to predict the observational properties of failed SN. Even though the details of the detectable signatures from this phenomenon are not tightly constrained, there is a general consensus that for red supergiants this mechanism will produce a low luminosity outburst, in the order of $10^{39}$ erg/s, followed by a plateau lasting several months. The ejected mass ranges from $10^{-3}M_\odot$ to several $M_\odot$ with low expansion velocities and kinetic energies, $v_e\sim 100$~km/s and $K_e\lesssim 10^{48}$~erg, which leads to dust formation surrounding the remnant \citep[see, e.g.,][]{Lovegrove2013,Kochanek2014fSNdust,Lovegrove2017,Fernandez2018,Ivanov2021,Antoni2023}.

The Large Binocular Telescope (LBT) has a dedicated program to search for failed SN candidates and place constrains on their rates by monitoring 27 nearby galaxies since 2008. The search has identified one promising candidate in the galaxy NGC~6946, \bh1 \citep{Gerke2015}. The discovery of one candidate in the survey implies a failed SN rate consistent with theoretical expectations \citep{Neustadt2021}. 

The progenitor, outburst and later evolution of \bh1 after its discovery \citep{Adams2017BH1,Basinger2021,Kochanek2024} closely resemble those predicted by failed SN models. The progenitor was a red supergiant of $\sim$25$M_\odot$ and $\sim$10$^{5.6}L_\odot$ and the optical weak ($\sim$10$^6L_\odot$) outburst detected in 2009 was followed by its disappearance in the optical over the next 3 to 11 months. Fifteen years after the outburst the source remains optically invisible but with mid-infrared (MIR) emission that is roughly 10 times fainter than the progenitor.

\cite{De2026Science} searched the Near-Earth Object Wide-Field Infrared Survey Explorer (NEOWISE) MIR sky survey from 2009 to 2022 for possible failed SNe in the Andromeda (M31) and Triangulum (M33) galaxies and found a compelling candidate in the former, M31-2014-DS1. The progenitor was a yellow supergiant star of $\sim$13$M_\odot$ and $\sim$10$^5L_\odot$ surrounded by a dust shell. M31-2014-DS1 had a MIR transient in 2014 in which its MIR flux increased by 50\% during 2 years but an optical outburst was not detected, with an upper limit of $10^5L_\odot$ over less than 180 days \citep{De2026Science}. Ten years after the outburst, the remnant is only 7\% of the progenitor luminosity and dominated by MIR dust emission \citep{De2026}. \cite{Nakanishi2026} searched for neutrino emission during the transient with Super-Kamiokande, but detected no neutrinos, albeit with flux limits that are not constraining for core collapse models.

These events are clearly not SNe, but stellar mergers (sometimes called luminous red novae, LRNe) are argued to have similar properties, like an optical transient followed by dust formation and the obscuration of the source \citep[see, e.g.,][for a review]{Kaminski2026}. However, even if the optical transients may be similar, the late time properties of the two scenarios must be significantly different. For a failed SN, the late time emission is due to residual accretion onto the newly formed black hole. Even at the Eddington limit, this luminosity is at most comparable to that of the progenitor, and the accretion rate is expected to steadily drop \citep[see, e.g.,][]{Lovegrove2013,Fernandez2018,Faran2026}. In contrast, a stellar merger forms an over-inflated, more massive star which should fade as its envelope returns to equilibrium but to a final luminosity significantly greater than that of the progenitors. \cite{Beasor2024} and \cite{Beasor2026} try to explain the low luminosity of N6946-BH1 and M31-2014-DS1 as a consequence of viewing the systems through a nearly edge-on, dusty disk but interpreting them with spherical dust models, but \cite{Kochanek2024dusty} had already shown that this model is incapable of decreasing the inferred luminosity more than a factor of 3 and has significant difficulty explaining the absence of optical emission in the observed systems even at extremely high (visual $\tau\sim10^3$) optical depths. 

In this paper we combine and reanalyze the JWST data on \bh1. The data and the analysis methods are described in Section~\ref{sec:data}. In Section~\ref{sec:neigh} we discuss the source identification, in particular finding that the bluer near-infrared sources are unrelated red giants. In Section~\ref{sec:n6946} we analyze the combined spectral energy distribution of \bh1 and the optical evolution since it was detected. In Section~\ref{sec:mergers} we compile Galactic and extragalactic examples of stellar mergers and carry out a systematic comparison of their properties with those of the two failed SN candidates in Section~\ref{sec:vs}. The conclusions are presented in Section~\ref{sec:concl}.

\section{Data and analysis}\label{sec:data}

Using MAST we obtained the JWST and HST data listed in Table~\ref{tab:obs}. These come from JWST programs 2896 (PI: Kochanek) and 3773 (PI: Beasor)\footnote{Both programs have been reprocessed by STScI with the Data Processing software version 2025\_2.} plus HST program 11229 (PI: Meixner). The HST data show the progenitor several years before the 2009 transient and the JWST data were obtained roughly 14 years post-transient. Additionally, the ongoing monitoring program with the LBT has seven new measurements of \bh1 in the $R$ band since the light curve in \cite{Kochanek2024}. The LBT observations were acquired with the LBC-Red camera \citep{Giallongo2008,Speziali2008} with a magnitude limit of approximately 25.9 mag. The details of the LBT observing program can be found in \cite{Gerke2015}.

\begin{table*}
    \renewcommand{\arraystretch}{1.5}
    \caption{Summary of the observations used for source identification and photometry extraction.}
    \label{tab:obs}
    \centering
    \begin{tabular}{ccccr@{ $\times$ }lc}
    \hline
    Telescope & Instrument & Filter & Date (yyyy/mm/dd) & Texp (s) & Nexp & Program \# (PI) \\
    \hline
    HST & WFPC2 & F814W & 2007/07/08 & 400.000 & 4 & 11229 (Meixner) \\
    \hline
    \multirow{4}{*}{JWST} & \multirow{4}{*}{NIRCam} & F115W & \multirow{4}{*}{2023/08/25} & 1052.203 & 4 & \multirow{4}{*}{3773 (Beasor)}\\
                               &  & F182M &  & 1052.203 & 4 & \\
                               &  & F250M &  & 526.102 & 4 & \\
                               &  & F360M &  & 526.102 & 4 & \\
    \hline
    \multirow{4}{*}{JWST} & \multirow{4}{*}{MIRI} & F560W & \multirow{4}{*}{2023/08/25} & 58.276 & 4 & \multirow{4}{*}{3773 (Beasor)}\\
                               &  & F770W &  & 30.525 & 4 & \\
                               &  & F1000W &  & 30.525 & 4 & \\
                               &  & F2100W &  & 30.525 & 4 & \\
    \hline
    \multirow{3}{*}{JWST} & \multirow{3}{*}{MIRI} & F560W & \multirow{3}{*}{2023/09/26} & 21.632 & 4 & \multirow{3}{*}{2896 (Kochanek)}\\
                               &  & F1000W &  & 21.632 & 4 & \\
                               &  & F2100W &  & 131.523 & 4 & \\
    \hline
    \end{tabular}
\end{table*}

The JWST data are processed with standard \texttt{DOLPHOT} routines \citep{Dolphin2000,Dolphin2016,Weisz2024}. We use Stage 2 images to extract the photometry using the NIRCam F115W and MIRI F560W Stage 3 images as reference frames for the NIRCam and MIRI images, respectively. The LBT data are processed as described in \cite{Gerke2015}, \cite{Adams2017search}, and \cite{Neustadt2021}.

To properly identify \bh1 in the JWST images and to extract the photometry in the crowded LBT images, we use the \texttt{ISIS} image subtraction package \citep{Alard1998,Alard2000}. This software allows to align and subtract images from each other to both accurately determine the source position and to reveal flux differences between epochs or wavelengths.

The spectral energy distributions (SEDs) are fit using \texttt{DUSTY} \citep{Ivezic1997,Ivezic1999,Elitzur2001} inside a Markov Chain Monte Carlo (MCMC) driver. For the central source we use either \cite{Castelli2003} or MARCS \citep{Gustafsson2008} model stellar atmospheres with variable temperature $T_\ast$ and luminosity $L$. We use the \cite{Castelli2003} stellar atmospheres for a broad range of temperatures, from 3500 to 39000~K, with $\log g=4$, $[Fe/H]=0$, and a microturbulence of $\xi=2$~km/s, while the MARCS models are used for lower temperatures, from 2600 to 4000~K, with $\log g=0$, $[Fe/H]=0$, and $\xi=2$~km/s. For our SED fitting purposes, the only relevant parameter of the stellar atmosphere models is the temperature. The surrounding material is modeled by a spherically symmetric shell of dust extending from $R_\text{in}$ to $R_\text{out}$ with density $\rho\propto r^{-2}$. The variable dust parameters are the temperature at $R_\text{in}$ $(T_d$), the visual optical depth ($\tau_V$), and the shell thickness ($R_\text{out}/R_\text{in}$). We use either \cite{Draine1984} graphitic or silicate dusts with a \cite{Mathis1977} grain size distribution, $dn/da \propto a^{-3.5}$, initially spanning sizes from $a_\text{min}$ to $a_\text{max}$ with 0.005 and 0.25 \micron, respectively, as default values. If the SED fit is unsatisfactory, we use a grid of models changing $a_\text{max}$ to find a best value. Due to likely systematic errors in the photometry and in the models, we adopt a minimum flux error of 10\% for the SED model. 

\section{Neighbors of N6946-BH1}\label{sec:neigh}

\begin{figure}
\centering
\includegraphics[width=0.47\textwidth]{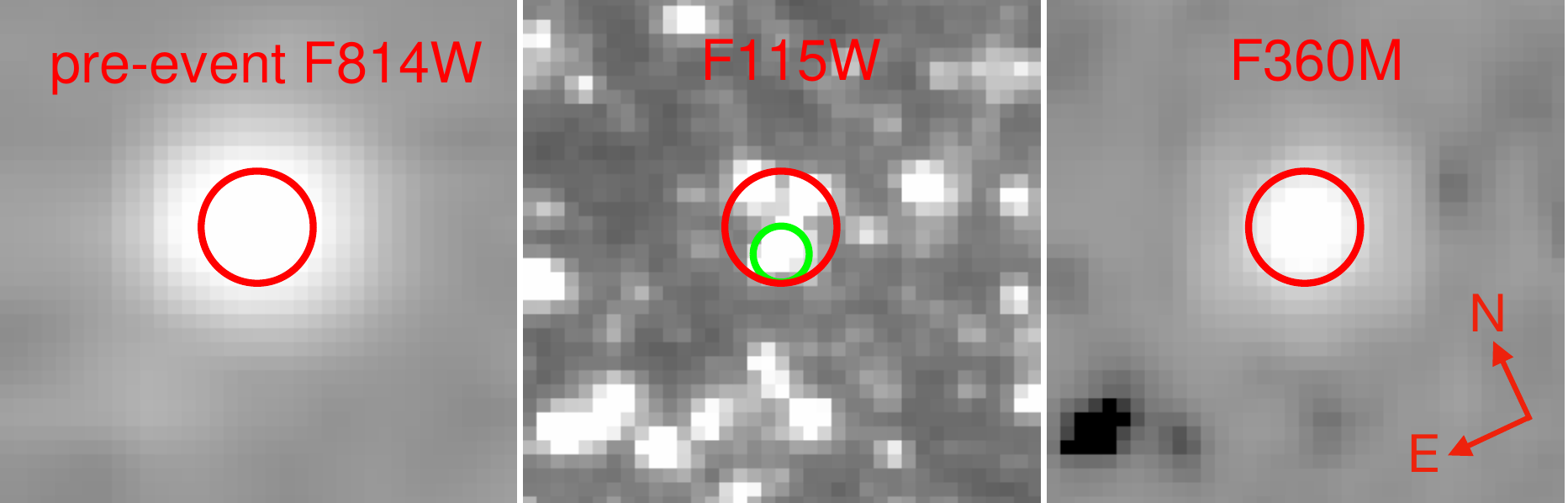}
\caption{The HST F814W pre-event image of \bh1 (left) and the JWST NIRCam F115W (center) and F360M (right) images 14 years after the event. The red 0\farcs124 radius circle is centered at the source position of the F814W pre-event image and the green 0\farcs062 radius circle in the F115W image is the source position adopted by \cite{Beasor2024}. The circle sizes are greater than the $\sim$0\farcs01 rms accuracy of the alignment between the images (see text). The panels are 2\farcs33 across and the orientation is shown in the lower right corner of the F360M panel.
  }
\label{fig:source}
\end{figure}

Examining the NIRCam images in Figure~3 of \cite{Beasor2024}, there are several sources near the source position at the shorter higher-resolution bands F115W and F182M, while the source in the F360M image is slightly north of the position used for \bh1 by \cite{Beasor2024}. Since luminous dusty stars are rare, it seems likely that the emission at the F360M band and longer wavelengths must be the counterpart to \bh1. For this reason, it is necessary to revisit the source identification in the NIRCam images. 

We used \texttt{ISIS} to align the pre-event HST WFPC2 F814W
image and the four NIRCam images.  We extracted a $321\times321$ pixel
 region roughly centered on the
source from the Stage 3 F115W and F182M images, and linearly interpolated the Stage 3 F250M and F360M images
to the pixel scale of the shorter wavelength images using IRAF {\tt magnify}. We did 
the same with the F814W image after a $90^\circ$ rotation.  Using the F115W image as
the astrometry reference, we used \texttt{ISIS} to interpolate the other images to the frame of
the F115W image.  It matched 404, 253, 167 and 20 stars for the F182M, F250M, F360M and F814W 
images with rms residuals of $0.06$ ($0\farcs002$), $0.28$ ($0\farcs009$), $0.29$ 
($0\farcs009$) and $0.36$ ($0\farcs011$) pixels.  We used \texttt{ISIS} to subtract the F182M image scaled
in flux and point spread function (PSF) structure from each of the other images.  This approach should leave
subtracted F814W and F360M images that are completely dominated by the pre-event star for
the former and the dusty source in the latter.  

\begin{figure*}
\centering
\includegraphics[width=\textwidth]{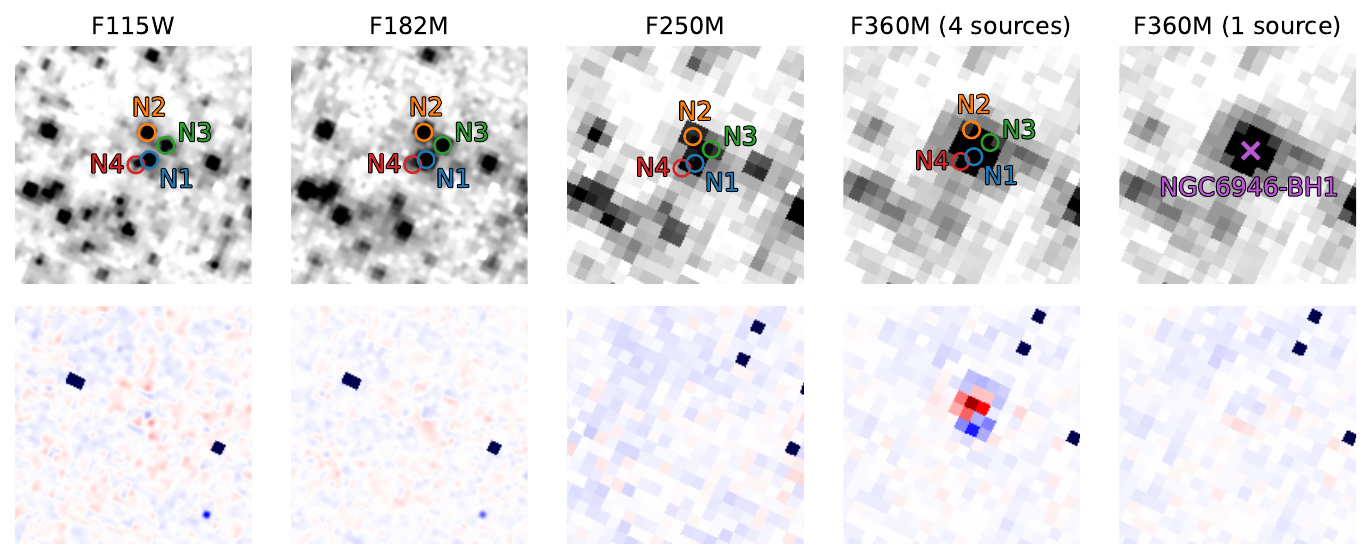}
\caption{Stage 2 images in each NIRCam band of Program 3773 (PI: Beasor). The blue, orange, green, and red circles are N1, N2, N3, and N4, respectively, and the purple cross in the right-most panel is placed at the \texttt{DOLPHOT} position of \bh1 in the F360M image. The second row of images are the residuals from the PSF fitting assuming only the presence of the 4 neighbors for the first four columns and assuming only the presence of \bh1 for the F360M image in last column. The orientation of the panels is north up, east left and they are 1\farcs25$\times$1\farcs25 in size.
  }
\label{fig:allimages}
\end{figure*}

\begin{table}
    \renewcommand{\arraystretch}{1.5}
    \caption{AB magnitudes of the four neighbors of \bh1.}
    \label{tab:neigmag}
    \centering
    \begin{tabular}{cr@{$\pm$}lr@{$\pm$}lr@{$\pm$}l}
    \hline
    Source & \multicolumn{2}{c}{F115W} & \multicolumn{2}{c}{F182M} & \multicolumn{2}{c}{F250M} \\
    \hline
    N1 & 25.453 & 0.018 & 24.853 & 0.015 & 25.364 & 0.019 \\
    N2 & 25.544 & 0.017 & 24.976 & 0.014 & 25.680 & 0.026 \\
    N3 & 25.905 & 0.017 & 25.366 & 0.019 & 26.106 & 0.037 \\
    N4 & 27.068 & 0.091 & 26.868 & 0.081 & 28.347 & 0.262 \\
    \hline
    \end{tabular}
\end{table}

Figure~\ref{fig:source} shows the results.  The central panel shows the F115W image
with the smaller, two pixel (0\farcs062) green circle marking the source position adopted
by \cite{Beasor2024}, which corresponds to the red giant we label N1 below (see Figure~\ref{fig:allimages}).  This position corresponds to the lower of the three central
sources (N1-N3) seen in the image. The left panel shows the subtracted, pre-event HST F814W
image and the right panel shows the subtracted JWST F360M image.  The larger,
four pixel (0\farcs124) circle in each panel is centered on the location of the source in the F360M
image.  We see that this exactly corresponds to the location of the star seen
prior to the event and does not correspond to the \cite{Beasor2024} source.  
This can also be seen in Figure~3 of \cite{Beasor2024} where the F360M source
is slightly above their adopted source position.  The centroid of the F814W
source is offset from the centroid of the F360M source by $0.60$ pixels (0\farcs018),
far less than the FWHM of point sources in either image ($0\farcs12$ for F360M,
and determined by sampling and the $0\farcs10$ pixel scale of the wide field
channels of WFPC2). There are hints in the F115W image that the lower source
in the triangle is not as point-like as the other two (our later \texttt{DOLPHOT} analysis confirms the presence of a fourth source, N4, next to N1), suggesting that the 
counterpart to the F360M source is a fainter object inside the triangle of brighter sources.
Similarly, in the subtracted F250M image, there are hints of a source at the
position of the F360M source, suggestive of some dust emission at this shorter
wavelength as well.

\texttt{DOLPHOT} finds four sources in the neighborhood using the Stage 3 NIRCam F115W image for its reference frame. Three of the sources are easily seen in Figure~\ref{fig:source} and a fourth, dimmer one is located to the east of the \cite{Beasor2024} source (see Figure~\ref{fig:allimages}). These four sources are distinguishable in the F182M and F250M filters as well, whereas in the Stage 3 F360M image \texttt{DOLPHOT} identifies a single source, centered where \texttt{ISIS} finds the dusty source. The AB magnitudes in these three bluest filters of the four sources are extracted using \texttt{DOLPHOT}'s standard PSF photometry extraction and presented in Table~\ref{tab:neigmag}. The reference frame for the photometry is the Stage 3 NIRCam F115W image and the rms residuals of the alignment for the 12 Stage 2 NIRCam images span 0.06 to 0.08 pixels or, equivalently, 0\farcs002 to 0\farcs006. The N1 neighbor (blue circle in Figure~\ref{fig:allimages}) is the source identified by \cite{Beasor2024} and the sources N2, N3, and N4 are the orange, green, and red circles, respectively.

\begin{table}
    \renewcommand{\arraystretch}{1.5}
    \caption{Fit parameters and 95\% confidence intervals for the four neighbors of \bh1.}
    \label{tab:neigparam}
    \centering
    \begin{tabular}{cSr@{}lr@{}l}
    \hline
    Source & {$\chi_\nu^2$} & \multicolumn{2}{c}{$T_\ast$ (K)} & \multicolumn{2}{c}{$\log(L/L_\odot)$} \\
    \hline
    N1 & 1.33 & 4000&$\pm500$ & 3.40&$_{-0.05}^{+0.09}$ \\
    N2 & 1.00 & 4000&$_{-500}^{+700}$ & 3.33&$_{-0.05}^{+0.13}$ \\
    N3 & 1.52 & 4000&$_{-500}^{+700}$ & 3.18&$_{-0.06}^{+0.14}$ \\
    N4 & 11.42 & 5000&$\pm1400$ & 2.8&$\pm0.2$ \\
    \hline
    \end{tabular}
\end{table}

\begin{figure}
\centering
\includegraphics[width=0.47\textwidth]{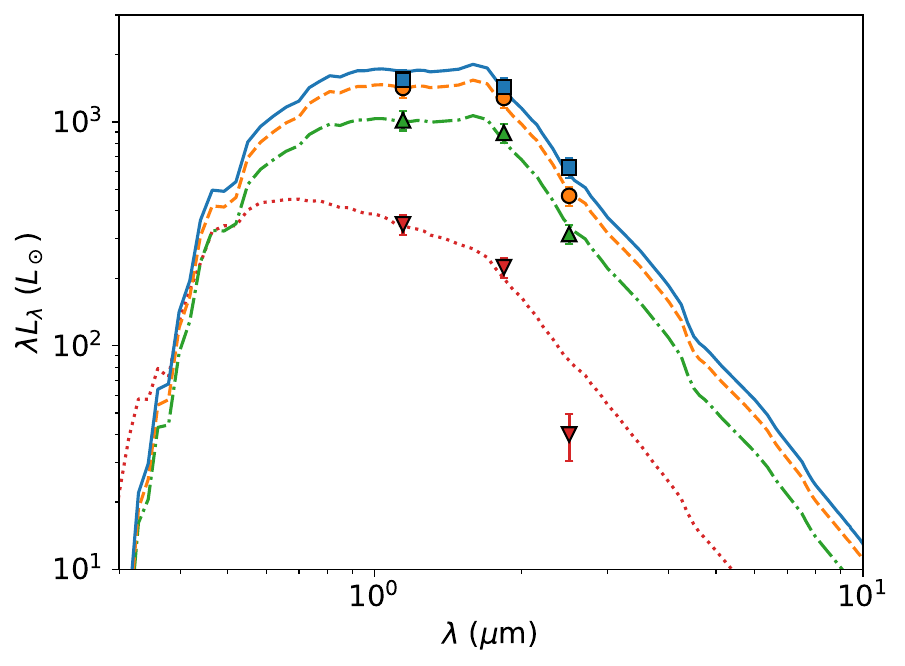}
\caption{The data (points) and best fit SED models (curves) are shown for N1 (blue squares and solid line), N2 (orange circles and dashed line), N3 (green upward triangles and dot-dashed line) and N4 (red downward triangles and dotted line).
  }
\label{fig:SEDneig}
\end{figure}

\begin{figure*}
\centering
\includegraphics[width=\textwidth]{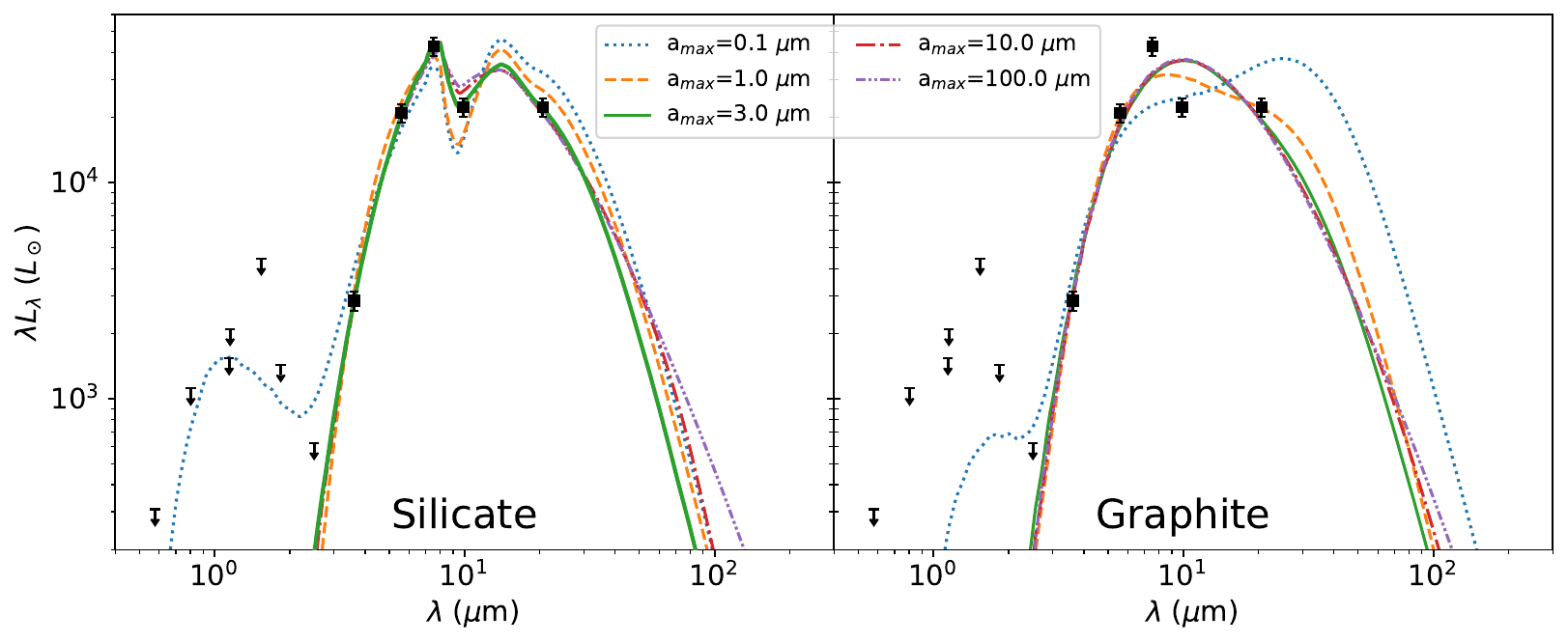}
\caption{SED fits to \bh1 with silicate (left panel) and graphitic (right panel) dusts. The maximum grain size is varied from 0.1 to 100 \micron\ and the best-fit model is the solid green line in the silicate panel.
  }
\label{fig:dustmod}
\end{figure*}

Adopting a distance to NGC~6946 of 7.7~Mpc \citep{Anand2018} and a total extinction of $E(B-V)=0.303$ \citep{Schlafly2011}, as in \cite{Basinger2021} and \cite{Kochanek2024}, we fit the SEDs of the neighbors with \texttt{DUSTY} using dust-free \cite{Castelli2003} stellar model atmospheres. We include no dust because we see no evidence for longer wavelength emission beyond that expected from stellar photospheres. Figure~\ref{fig:SEDneig} shows the best model fits with the same color coding as in Figure~\ref{fig:allimages} and the fit parameters with 95\% confidence intervals are presented in Table~\ref{tab:neigparam}. Although the uncertainties in the parameters are somewhat large, especially for $T_\ast$, due to the lack of measurements in the blue part of their SEDs (the sources are confused in HST optical images), the four sources are consistent with red giants in NGC~6946. Stars with these luminosities and temperatures in PARSEC \citep{2012MNRAS.427..127B,2014MNRAS.444.2525C,2015MNRAS.452.1068C,2014MNRAS.445.4287T,2017ApJ...835...77M,2019MNRAS.485.5666P,2020MNRAS.498.3283P} isochrones are predicted to have mass loss rates of $\sim10^{-8}$~$M_\odot$/year, consistent with the lack of significant dust emission.

\section{N6946-BH1}\label{sec:n6946}

In Section~\ref{sec:neigh}, we found that the progenitor and remnant positions of \bh1 do not correspond to any of the 4 neighbors (N1-N4) detected in the NIRCam F115W, F182M, and F250M bands. Additionally, the fluxes at wavelengths longer than the F360M band are dominated by the remnant emission of \bh1. In this section we characterize the SED of \bh1 using different dust models and infer its ejecta properties. We also update the $R$-band light curve since the start of LBT monitoring campaign until September 29, 2025.

\subsection{SED fit}\label{subsec:SEDbh1}

Given that the emission in the NIRCam F115W, F182M, and F250M filters is dominated by the four red giants, we can only place upper limits on the near-infrared flux of \bh1. We conservatively use the magnitudes of the brightest neighbor, N1, for the limits. If the remnant emission was brighter than N1, we would have easily seen it in the residual images of Figure~\ref{fig:allimages}. We merge the MIRI data from Programs 2896 and 3773 and analyzed them jointly with \texttt{DOLPHOT}. We use the MIRI F560W Stage 3 image from Program 3773 as the reference frame for the 8 F560W, 4 F770W, 8 F1000W, and 8 F2100W images. We use the F360M photometry at the position of the purple cross in Figure~\ref{fig:allimages}. The AB magnitudes for each band are presented in Table~\ref{tab:bh1mags}. The extracted fluxes in F1000W and F2100W agree well with the values reported in \cite{Beasor2024} and \cite{Kochanek2024} with only $\sim$5\% differences, which might be expected due to the different extraction methods. However, the other fluxes have larger differences. The F360M flux in \cite{Beasor2024} is 26\% smaller than the value reported here, while the F560W and F770W fluxes are 38\% and 32\% greater, respectively. The F560W flux in \cite{Kochanek2024} is 15\% smaller. The data analyzed here were reduced with the Data Processing software 2025\_2, a later version than the 2023\_2a used for the data in \cite{Kochanek2024} and \cite{Beasor2024}, and we see similar flux differences between the two versions of the STScI source catalog of Program 2896.

\begin{table}
    \renewcommand{\arraystretch}{1.5}
    \caption{Magnitudes and fluxes of \bh1 along with their uncertainties and the number of images used for each band, N$_\text{img}$.}
    \label{tab:bh1mags}
    \centering
    \begin{tabular}{ccr@{$\pm$}lr@{$\pm$}lc}
    \hline
    Instrument & Band & \multicolumn{2}{c}{AB magnitude} & \multicolumn{2}{c}{Flux ($\mu$Jy)} & N$_\text{img}$ \\
    \hline
    NIRCam & F360M & 23.281 & 0.005 & 1.768 & 0.008 & 4\\
    MIRI & F560W & 20.611 & 0.004 & 20.68 & 0.08 & 8 \\
    MIRI & F770W & 19.510 & 0.003 & 57.02 & 0.16 & 4 \\
    MIRI & F1000W & 19.913 & 0.005 & 39.34 & 0.18 & 8 \\
    MIRI & F2100W & 19.108 & 0.014 & 82.6 & 1.1 & 8 \\
    \hline
    \end{tabular}
\end{table}

We estimated the fractional contamination to the fluxes reported in Table~\ref{tab:bh1mags} from the N1-N4 neighbors to be 9.7\%, 0.62\%, 0.13\%, 0.13\%, and 0.018\% for the F360M, F560W, F770W, F1000W, and F2100W bands, respectively, by extrapolating the best fit dust-free stellar atmosphere models of Section~\ref{sec:neigh}.  These are less than our default 10\% error budget for the SED fitting.

To model the SED of \bh1, we use the magnitudes from Table~\ref{tab:bh1mags}, the upper limits from the N1 neighbor for the NIRCam F115W, F182M, and F250M bands and additional upper limits from the HST WFC3 F606W, F814W, F110W, and F160W bands from \cite{Kochanek2024}. We adopt the same distance and extinction as for the neighbors in Section~\ref{sec:neigh}. We vary the source luminosity, $L$, the dust temperature, and the optical depth. The underlying source has a fixed temperature of 10000~K, modeled with a \cite{Castelli2003} stellar atmosphere because the source temperature has little effect on the MIR part of the SED, as shown by \cite{Kochanek2024}. The dust thickness is fixed to $R_\text{out}=2R_\text{in}$. Changing it to $R_\text{out}=4R_\text{in}$ does not significantly improve the fit. We try both \cite{Draine1984} graphitic and silicate dusts spanning sizes from $a_\text{min}$=0.005 \micron\ to $a_\text{max}$ with $a_\text{max}$ ranging from 0.1 to 100 \micron. We run 1000 long MCMC chains on the converged models for each value of $a_\text{max}$.

\begin{figure}
\centering
\includegraphics[width=0.47\textwidth]{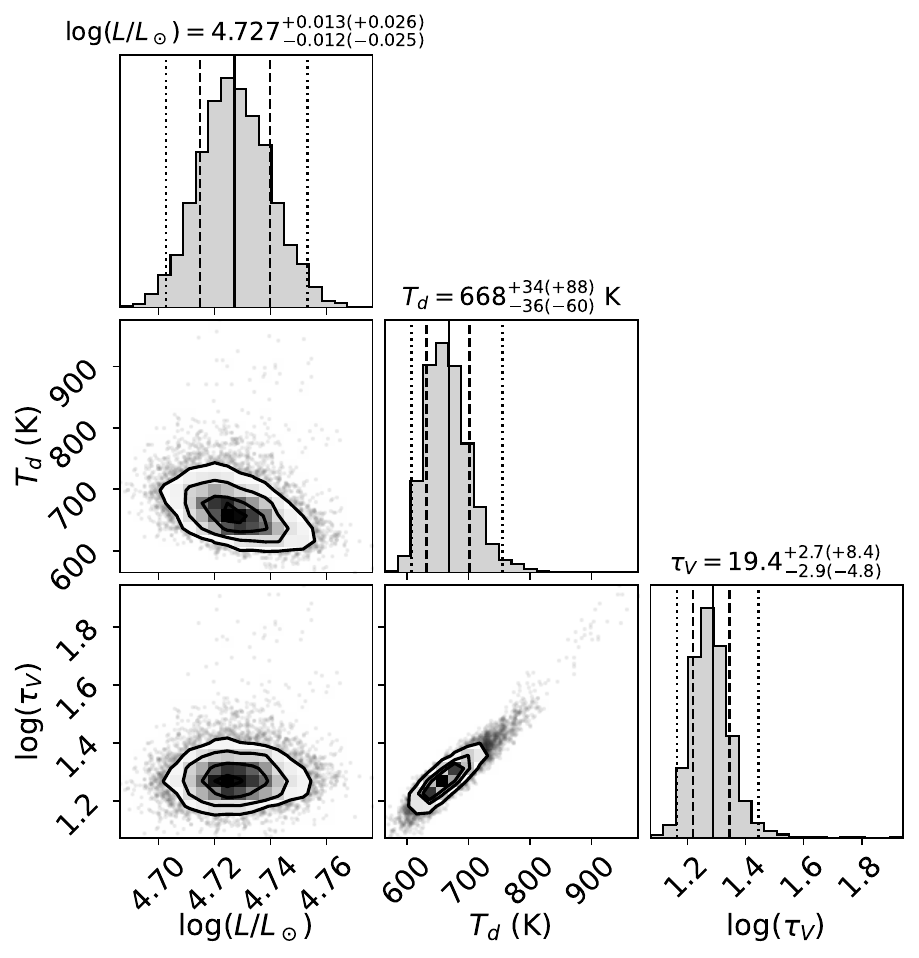}
\caption{MCMC corner plot \citep[using \texttt{corner.py},][]{corner} of the \bh1 SED fit parameters with a silicate dust with $a_\text{max}$=3.0~\micron. The optical depth is plotted in logarithm scale for visualization purposes. The solid vertical lines are placed at the best-fit values and the dashed (dotted) lines are placed at the 68\% (95\%) confidence levels. The 2D contours are drawn at 11.8\%, 39.3\%, 67.5\%, and 86.4\% confidence levels.
  }
\label{fig:mcmc}
\end{figure}

Figure~\ref{fig:dustmod} shows that graphitic dusts are unable to reproduce the $\sim$10 \micron\ feature for any maximum grain size. The data favors a silicate dust with a maximum grain size of 3 \micron, which is the solid green line on the left panel of Figure~\ref{fig:dustmod}. We run a longer MCMC chain (10000 steps) for the silicate model with $a_\text{max}$=3~\micron\ to infer the luminosity, dust temperature and optical depth confidence intervals. The best model fit has a $\chi^2_\nu$=0.15 and the 68\% (95\%) confidence parameter estimates are $\log(L/L_\odot)$=$4.727_{-0.012(-0.025)}^{+0.013(+0.026)}$, $T_d$=$668_{-36(-60)}^{+34(+88)}$~K and $\tau_V$=$19.4_{-2.9(-4.8)}^{+2.7(+8.4)}$. Figure~\ref{fig:mcmc} shows the MCMC corner plot for the $a_\text{max}$=3~\micron\ case. There is a weak correlation between dust temperature and optical depth. Although we have not varied the maximum grain size and the shell thickness inside the MCMC due to the limited number of data points, the derived luminosities and temperatures are little different. In particular, the estimates of the remnant luminosity for the other $a_\text{max}$ models are all within 13\% of the estimate from the best fit, even though the fits are significantly poorer.

\begin{figure}
\centering
\includegraphics[width=0.47\textwidth]{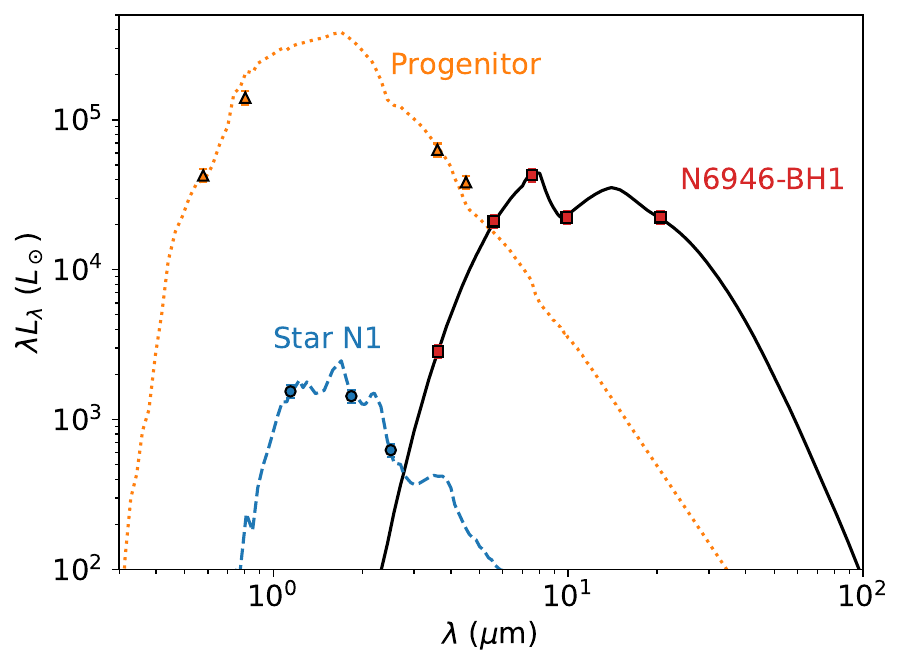}
\caption{Spectral energy distributions for \bh1 (red squares and black solid line), its progenitor (orange triangles and dotted line) and the neighboring star N1 (blue circles and dashed line) which was the source position for \bh1 used by \cite{Beasor2024}.
  }
\label{fig:SEDprogremn}
\end{figure}

\begin{figure*}
\centering
\includegraphics[width=\textwidth]{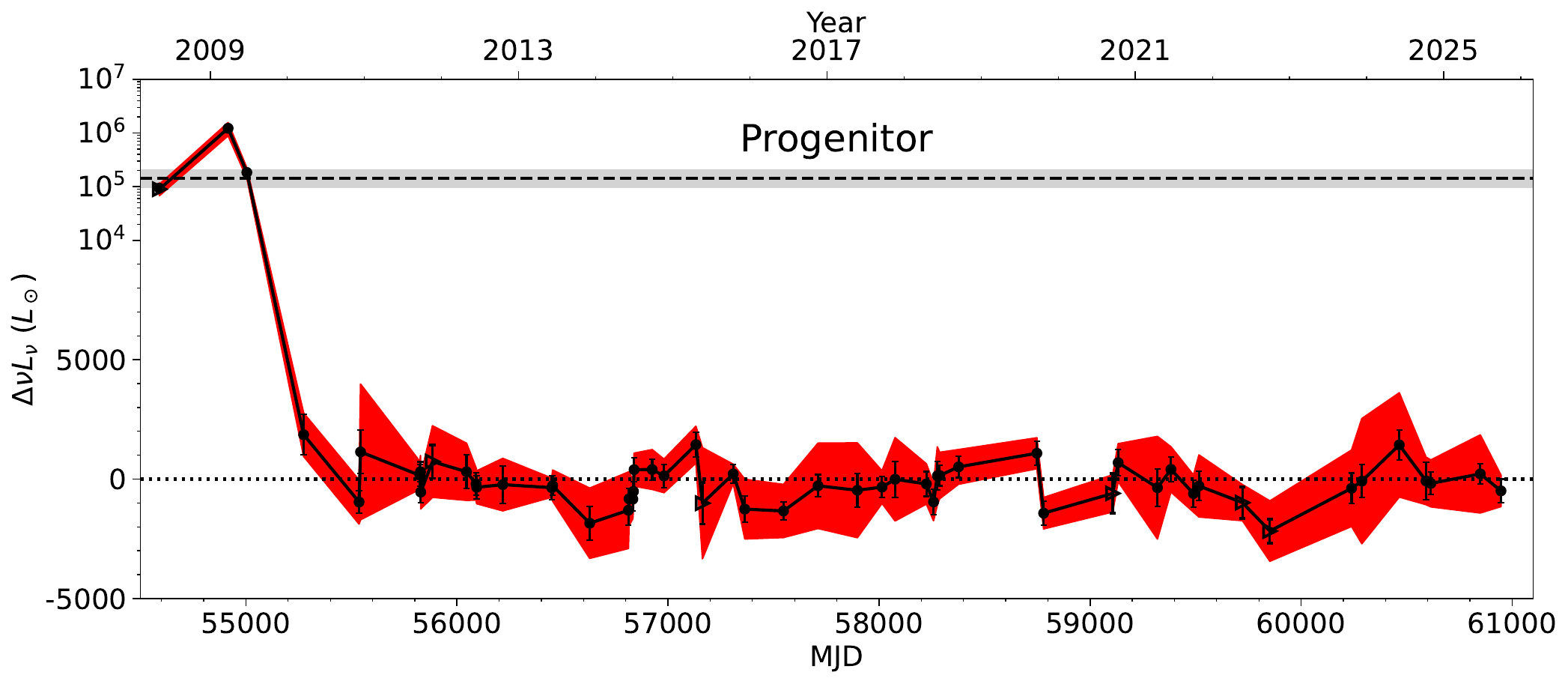}
\caption{LBT $R$-band light curve of \bh1. The dots (open triangles) are measurement taken under good (bad) observing conditions. The error bars are the formal \texttt{ISIS} light curve uncertainties while the red shaded region is the dispersion in the 25 light curves surrounding \bh1. The progenitor $R$-band luminosity is shown by the gray horizontal band and the dotted line is at zero luminosity for reference. The vertical axis is linear below $10^4L_\odot$ and logarithmic above that value.
  }
\label{fig:LBTlc}
\end{figure*} 

The estimated mass and kinetic energy of the ejecta for this silicate model are similar to \cite{Kochanek2024} models. Following Equations~1 and 2 of \cite{Kochanek2024} for the best model inner radius of $R_\text{in}=10^{15.42}$~cm, assuming the lower limit of $R_\text{out}=R_\text{in}$ and a ejecta velocity of $v_e\simeq R_\text{in}/\Delta t=60$~km/s using $\Delta t=14$~yr, we estimate $M_e\gtrsim 0.08 M_\odot$ and $K_e\gtrsim 3\times10^{-6}$ FOE (or, equivalently, $3\times10^{45}$ erg), which are smaller than expected for an SN. These are lower limits from using $R_\text{in}$ and ignoring $R_\text{out}$, and from allowing a significant contribution from small grains.

Figure~\ref{fig:SEDprogremn} shows the best fit SED models for \bh1, its progenitor, and the nearby star N1. For the progenitor we used the 2007 HST F606W and F814W and Spitzer 3.6 and 4.5 \micron\ measurements \citep{Adams2017BH1} and fit it with a MARCS stellar atmosphere without dust. The progenitor luminosity and temperature are $\log(L/L_\odot)$=$5.632_{-0.008(-0.018)}^{+0.012(+0.023)}$ and $T_\ast$=$3515_{-23(-45)}^{+14(+29)}$~K at 68\% (95\%) confidence levels, consistent with \cite{Kochanek2024}. Since the NIR measurements reported by \cite{Beasor2024} are associated to the neighbor star N1, the shape of the SED of \bh1 is now well described by a central source surrounded by a shell of silicate dust, without the need to invoke superwinds or polycyclic aromatic hydrocarbon (PAH) emission.

\subsection{LBT light curve}

We have continued to monitor \bh1 with the LBT and there are currently 56 epochs in $R$ band spanning from May 2008 to September 2025. The measurements until September 2022 are in \cite{Kochanek2024}. Here we update the light curve with the seven additional measurements presented in Table~\ref{tab:LBTlc}. The fluxes are extracted using \texttt{ISIS} difference imaging. The formal \texttt{ISIS} errors tend to be underestimates, so a grid of light curves extracted at 25 points placed around the source \citep[see][Figure~3]{Basinger2021} are used to estimate more realistic uncertainties. The updated light curve is shown in Figure~\ref{fig:LBTlc} where the progenitor $R$-band luminosity is estimated from the magnitudes from 2003 and 2005 presented in \cite{Adams2017BH1}. After the outburst, the $R$-band light curve luminosity is compatible with 0, with a sample standard deviation smaller than $10^3L_{\odot}$ and without either increasing or decreasing luminosity trends. A linear fit of the light curve since December 2015 gives a slope of $-2.1\pm24.5\, L_\odot/$yr, completely compatible with no changes.

\begin{table}
    \renewcommand{\arraystretch}{1.5}
    \caption{New $R$-band measurements of \bh1 from LBT.}
    \label{tab:LBTlc}
    \begin{center}
    \vspace{-1em}
    \begin{tabular}{crrrc}
    \hline
    MJD & $L$ ($L_\odot$) & $\epsilon$ ($L_\odot$) & $\sigma_L$ ($L_\odot$) & Flag \\
    \hline
     60239.26 & $-$380 & 630 & 1600 & 1 \\
     60288.08 &  $-$80 & 690 & 2620 & 1 \\
     60465.41 &   1430 & 620 & 2180 & 1 \\
     60592.12 &  $-$75 & 780 &  990 & 1 \\
     60615.11 & $-$180 & 470 &  980 & 1 \\
     60849.33 &    220 & 430 & 1630 & 1 \\
     60947.29 & $-$490 & 490 &  660 & 1 \\
    \hline
    \end{tabular}
    \end{center}
    
    {\footnotesize \textbf{Notes.} $L$ is the luminosity of the source, $\epsilon$ is the \texttt{ISIS} error estimate, and $\sigma_L$ is the luminosity dispersion of the 25 comparison grid points. These quantities are rounded to the nearest 10$L_\odot$. The flag value corresponds to good (1) or bad (0) observing conditions.}
\end{table}

\section{The properties of stellar mergers}\label{sec:mergers}

\cite{Beasor2024} and \cite{Beasor2026} discuss the hypothesis that \bh1 and M31-2014-DS1, respectively, could be stellar mergers instead of failed SNe. Here we compile the properties of Galactic and extragalactic stellar mergers, sometimes called luminous red novae, where both the progenitor and remnant were detected to compare their properties with those of \bh1 and M31-2014-DS1.

\begin{figure*}
\centering
\includegraphics[width=0.47\textwidth]{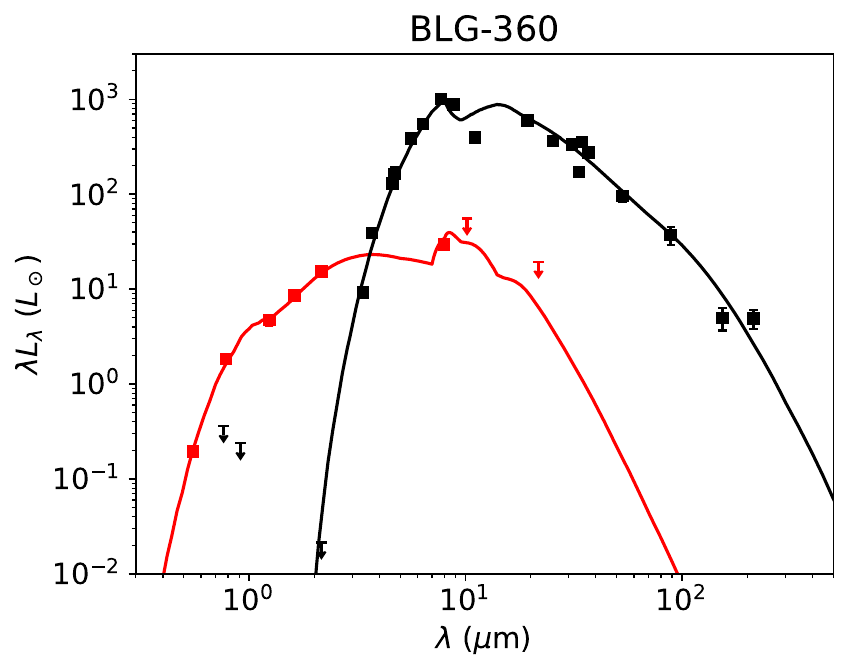}
\includegraphics[width=0.47\textwidth]{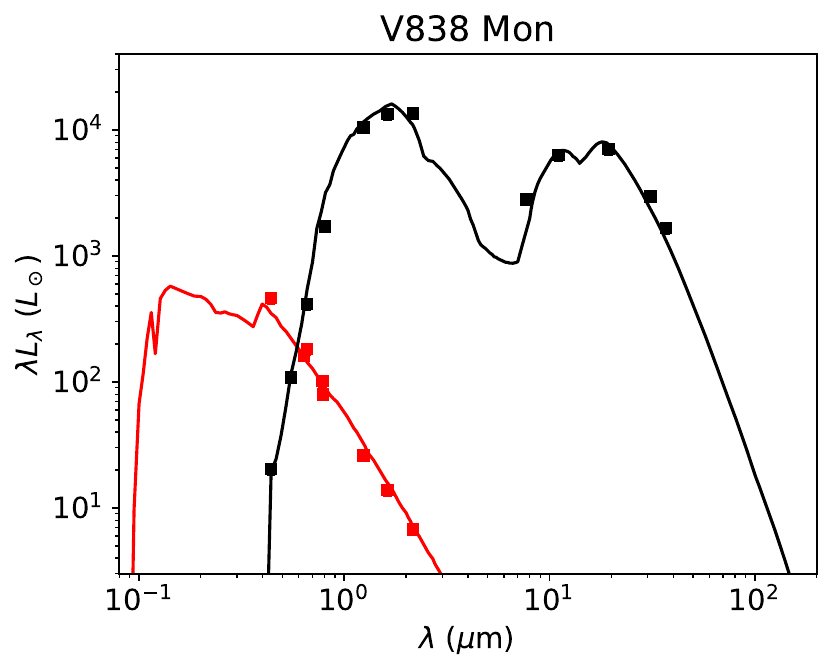}
\includegraphics[width=0.47\textwidth]{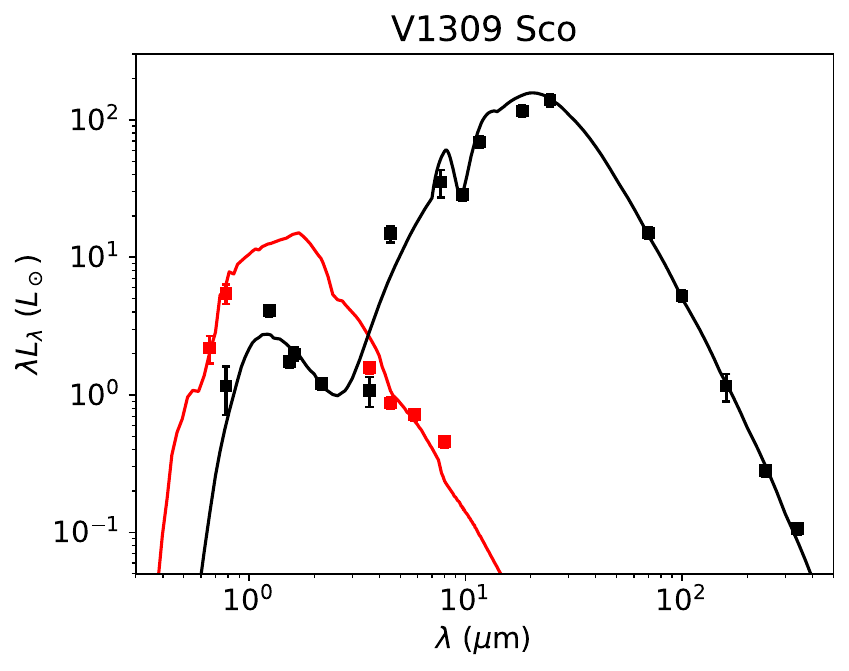}
\includegraphics[width=0.47\textwidth]{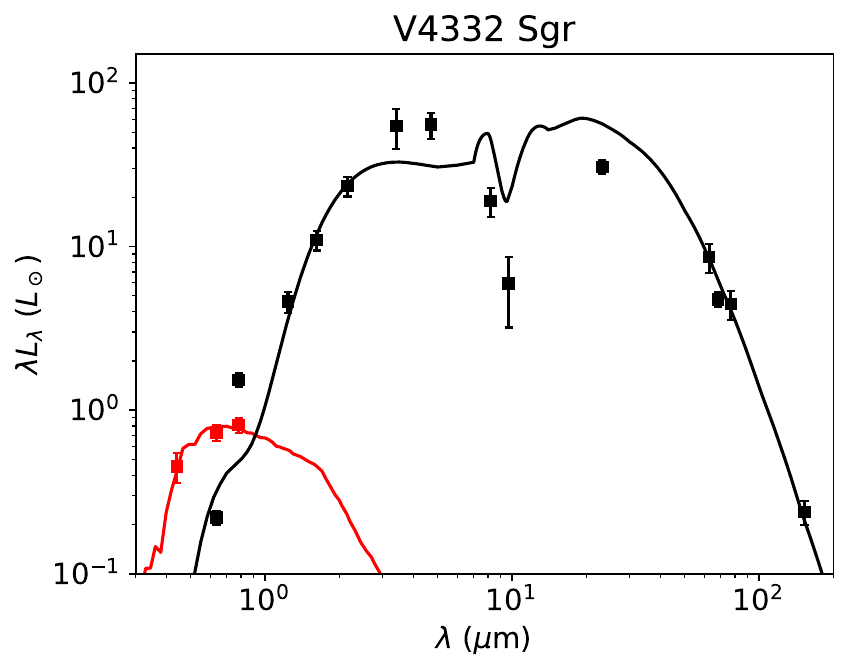}
\caption{
  SED fits to the progenitors (red) and remnants (black) of the Galactic stellar mergers BLG-360 (upper left), V838 Mon (upper right), V1309 Sco (lower left), and V4332 Sgr (lower right). 
  }
\label{fig:mergers}
\end{figure*}

\begin{table*}
    \renewcommand{\arraystretch}{1.5}
    \caption{Summary of best fit parameters for the Galactic stellar mergers progenitors and remnants.}
    \label{tab:summary}
    \centering
    \begin{tabular}{lc|c|SrSSrcc}
    \hline
    \multicolumn{2}{c|}{\multirow{2}{*}{Object}} & {Literature}&  \multicolumn{7}{c}{\texttt{DUSTY} best fit}\\
    & & {$\log (L/L_\odot)$} & {$\chi^2_\nu$}& {$T_\ast$ (K)} & {$\log (L/L_\odot)$} & {$\tau_V$} & {$T_d$ (K)} & $R_\text{out}/R_\text{in}$ & $a_\text{max}$ (\micron) \\
    \hline
    \multirow{2}{*}{BLG-360} & Progenitor & 1.86 & 53.5 & 9452 & 1.75 & 10.4 & 1231 & 2 (fixed) &  0.25 (fixed)\\
    & Remnant after 18 yrs & 3.22 & 12.1 & 25051 & 3.12 & 626.6 & 1501 & 3.61 & 10 (fixed)\\
    \hline
    \multirow{2}{*}{V838 Mon} & Progenitor & 2.74 & 5.8 & 14033 & 2.87 & {} & {} &  & \\
    & Remnant after 17 yrs & 4.56 & 15.3 & 3559 & 4.37 & 7.0 & 262 & 2.05 & 0.25 (fixed)\\
    \hline
    \multirow{2}{*}{V1309 Sco} & Progenitor & 0.93 & 43.0 & 3442 & 1.22 & {} & {} & & \\
    & Remnant after 4 yrs & 2.34 & 11.7 & 14654 & 2.32 & 17.8 & 460 & 1705 & 0.25 (fixed)\\
    \hline
    \multirow{2}{*}{V4332 Sgr} & Progenitor & & 0.8 & 5024 & 0.04 & {} & {} &  & \\
    & Remnant after 13 yrs & & 42.0 & 29315 & 2.13 & 6.0 & 1344 & 73.9 & 0.10 (fixed)\\
    \hline
    \end{tabular}
\end{table*}

\subsection{Galactic stellar mergers}\label{subsec:galmergers}

There are four known Galactic stellar mergers whose progenitors where observed before the outburst that led to the remnant detected today. We fit the available photometric measurements for both the progenitors and remnants with \texttt{DUSTY}.

\begin{table*}
    \renewcommand{\arraystretch}{1.5}
    \caption{Progenitor and remnant luminosities for extragalactic stellar mergers. The elapsed time, $\Delta t$, between the outburst and when the remnant luminosity was measured is also presented.}
    \label{tab:extramergers}
    \centering
    \begin{tabular}{lSlScl}
    \hline
    Object & {$\log (L_\text{prog}/L_\odot)$} & Reference & {$\log (L_\text{remn}/L_\odot)$} & {$\Delta t$} & Reference \\
    \hline
    AT~2015dl (M101-2015OT1) & 4.94 & \cite{Blagorodnova2017} & 5.57 & $\sim$500 days &  \cite{Blagorodnova2017}\\
    AT~2018bwo & 4.26 & \cite{Blagorodnova2021} & 5.17 & $\sim$6 years &  \cite{Karambelkar2026}\\
    AT~2019zhd & 3.00 & \cite{Pastorello2021I} & 4.00 & $\sim$3 years & \cite{Reguitti2026}\\
    AT~2020hat & 3.47 & \cite{Pastorello2021II} & 6.04 & $\sim$100 days & \cite{Pastorello2021II}\\
    AT~2021biy & 5.00 & \cite{Cai2022III} & 5.62 & $\sim$3 years & \cite{Karambelkar2026}\\
    AT~2021blu & 4.61 & \cite{Pastorello2023IV} & 5.52 & $\sim$3 years & \cite{Karambelkar2026}\\
    M31-LRN-20215 & 2.65 & \cite{Williams2015} & 3.66 & $\sim$10 years & \cite{Karambelkar2026}\\
    \hline
    \end{tabular}
\end{table*}

The OGLE-2002-BLG-360 (BLG-360 henceforth) outburst was identified in 2002 by the OGLE\footnote{https://ogle.astrouw.edu.pl/} Early Warning System and reached its maximum optical brightness in September 2003. We use the progenitor magnitudes and upper limits from \cite{Tylenda2013} that were acquired roughly between 1983 to 2001 to fit its SED while the remnant measurements are from 2021 and 2022 \citep{Steinmetz2025}. We adopt a distance of 4.09~kpc and a foreground extinction of $E(B-V)$=1.83 from \cite{Steinmetz2025}.

V838 Monocerotis (V838 Mon) is another Galactic example of a stellar merger that occurred in 2002 \citep{Brown2002}. The most recent photometric observations of the remnant are from 2019-2020 by \cite{Woodward2021} and, to characterize the progenitor, we use the same photometry as \cite{Tylenda2005V838Mon} which was acquired from 1994 to 2001. We adopt a distance of 6.1~kpc from \cite{Sparks2008} and a foreground extinction of $E(B-V)$=0.87 like \cite{Woodward2021}.

V1309 Scorpii (V1309 Sco) erupted in 2008 \citep{Nakano2008}. Spectroscopic observations by \cite{Mason2010} indicated that the object might be a stellar merger and, thanks to preexisting photometry from the OGLE project, the inspiral phase of the contact binary progenitor was detected \citep{Tylenda2011}. We use the broad-band photometry from 2012 to constrain the remnant SED and the photometry from 2007 for the progenitor (\citealt{McCollum2014} and \citealt{Tylenda2016}). The assumed distance of 3~kpc and the foreground extinction of $E(B-V)$=0.8 are from \cite{Tylenda2011}.

The oldest Galactic merger is V4332 Sagittarii (V4332 Sgr) whose outburst in 1994 \citep{Hayashi1994} was analyzed by \cite{Martini1999}. There are limited photometric data for the progenitor, and we use the magnitudes from the SuperCOSMOS and the USNO-B1.0 catalogs reported by \cite{Tylenda2005V4332Sgr}, which are built from scans of photometric plates mainly obtained in the 1970s. The remnant SED is modeled using the photometry from \cite{Kaminski2010} dating from 2005 to 2009. We adopt the same distance (1.8~kpc) and foreground extinction ($E(B-V)$=0.32) as \cite{Tylenda2005V4332Sgr} and \cite{Kaminski2010}.

A fifth Galactic candidate is CK Vulpeculae (CK Vul), also known as Nova 1670. Given that the outburst occurred in 1670 \citep[see][for the historic outburst observations]{Shara1985}, the progenitor cannot be characterized and we could not add it to our sample. Although this remnant is sometimes thought to be the result of a stellar merger \citep[see, e.g.][]{Kato2003,Kaminski2015,Eyres2018,Kaminski2021,Tylenda2024}, there are unique characteristics that differ from other LRNe like a brighter outburst with multiple peaks, high expansion velocities, a faint ($\sim0.9L_\odot$) and cold ($\sim40$~K) remnant, and peculiar element abundances
\citep[see, e.g.,][]{Hajduk2013,Kaminski2015,Evans2016,Banerjee2020}. It is also still fully obscured today, suggesting a need for ongoing dust formation.

We fit the systems as in Sections~\ref{sec:neigh} and \ref{subsec:SEDbh1}. We varied the progenitor luminosities and temperatures, and for BLG-360, the dust properties. The other three progenitors show no evidence for circumstellar dust. Since there are no measurements in the blue part of the SED of V838 Mon progenitor, we adopt a soft prior on the temperature of 6000$\pm$2000~K, which includes the range of effective temperatures reported by \cite{Tylenda2005V838Mon}.

For all the remnants and the BLG-360 progenitor we used \texttt{DUSTY} models with silicate dusts since they all show the $\sim$10 \micron\ feature often associated with silicates. By default the shell thickness, $R_\text{out}/R_\text{in}$, is fixed to 2 unless the fits are unsatisfactory, in which case it is treated as another free parameter in the model. If the goodness of fit is still poor, then the default maximum grain size of 0.25 \micron\ is changed manually, but kept fixed during each MCMC \texttt{DUSTY} run. 

The best fit models for the progenitors and remnants of each system are presented in Figure~\ref{fig:mergers} and the parameters and $\chi^2_\nu$ of the fits are summarized in Table~\ref{tab:summary}. Even if the thickness and maximum grain size are varied, the fits to the $\sim$10 \micron\ feature in BLG-360 and V4332 Sgr are not great. Better models likely require a 3D radiative transfer treatment and/or other choices for the dust composition. 

We are mainly interested in the SED fit as an interpolation model to estimate luminosities rather than as detailed dust study. However, Table~\ref{tab:summary} also lists previous luminosity estimates where available to confirm that only small deviations are present. The literature luminosity estimates are shifted to our adopted distances and isotropic emission is assumed where needed. The agreement is generally good ($\le 0.1$ dex) with the worst being 0.3 dex for the progenitor of V1309~Sco.

\subsection{Extragalactic stellar mergers}

\begin{figure*}
\centering
\includegraphics[width=\textwidth]{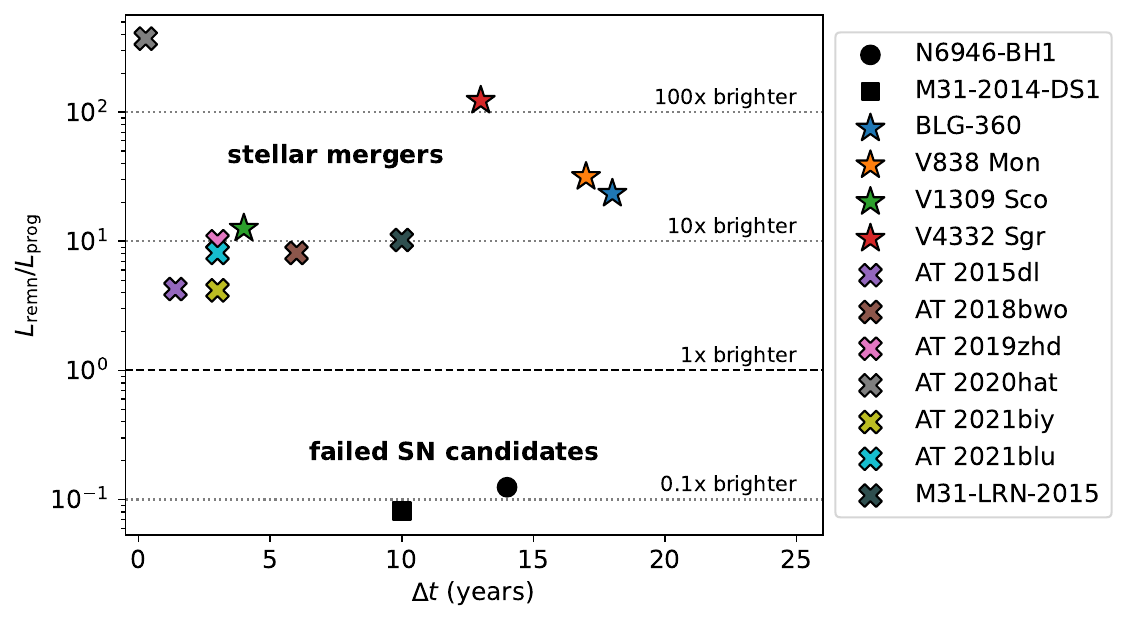}
\caption{Ratio of the remnant and progenitor luminosity as a function of the elapsed time since the outburst for \bh1 (black circle), M31-2014-DS1 (black square), Galactic stellar mergers (stars) and extragalactic stellar mergers (crosses).
  }
\label{fig:progremn}
\end{figure*}

There are over a dozen known extragalactic LRNe, but only seven of them have enough photometric measurements before and long after the outburst to estimate the luminosities of both the progenitor and the remnant. The limited number of available bands prevent us from modeling them with \texttt{DUSTY}, hence we adopt the values reported in the literature which are summarized in Table~\ref{tab:extramergers}.

We adopt the inferred bolometric luminosity of AT~2018bwo progenitor from the stellar spectral models fits of \cite{Blagorodnova2021}. The progenitor luminosities of AT~2019zhd, AT~2020hat, and M31-LRN-2015 are estimated by matching the extinction-corrected absolute magnitudes and colors from \cite{Pastorello2021I}, \cite{Pastorello2021II}, and \cite{Williams2015}, respectively, to PARSEC synthetic magnitudes in the corresponding filters. We generate isochrones with $\log(\text{Age}/yr)$ between 7.8 to 10.1 and $[M/H]$ ranging from $-2.0$ to $0.0$. The entries in the isochrones that have an absolute magnitude and color within the observed values and uncertainties are selected, and the median of their luminosities is reported as the progenitor luminosity in Table~\ref{tab:extramergers}. \cite{Reguitti2026} give a range of $\log (L/L_\odot)$ from 3.93 to 4.07 for the AT~2019zhd remnant, so we adopt the intermediate value of 4.00.

\section{Failed supernovae vs stellar mergers}\label{sec:vs}

M31-2014-DS1 \citep{De2026Science,De2026} is the only other failed supernova candidate. Its progenitor and the 2024 remnant luminosities were $\log (L/L_\odot)=$ 4.97 and 3.88, respectively. Figure~\ref{fig:progremn} compares the ratio between the remnant and the progenitor luminosities as a function of the elapsed time between the outburst and the remnant characterization for the stellar mergers from Section \ref{sec:mergers} and the failed supernovae candidates. \bh1 and M31-2014-DS1 are the only two systems with remnants dimmer than their progenitors. They fall in a very different part of the parameter space than the stellar mergers whose remnants are $\sim$10 to $\sim$100 times brighter than their progenitors even when the elapsed time since the outburst is comparable or longer than for the failed SN candidates. This is not surprising, since the envelopes of the stellar mergers are presumably in a state of Kelvin-Helmholtz contraction with times scales longer than decades. The evolution of $L_\text{remn}/L_\text{prog}$ for \bh1, M31-2014-DS1, BLG-360 and M31-LRN-2015 in \cite{De2026}, also show that \bh1 and M31-2014-DS1 remnants become fainter than their progenitors as opposed to BLG-360 and M31-LRN-2015, whose luminosities have a decreasing trend but remain well above their progenitor luminosities. The error bars for the luminosity ratios of the failed SN candidates are smaller than their markers and, more generally, the vertical scale of Figure~\ref{fig:progremn} is so large that even factor of 2 uncertainties would matter little.

Although it is likely that spherically symmetric dust models are not sufficient to describe such complex objects, \cite{Kochanek2024dusty} shows that the MIR luminosity of a source obscured by a dusty disk viewed edge-on can be underestimated only by up to a factor of $\sim$3 when an isotropic dust distribution is assumed even for sources obscured by a very high optical depth, exactly edge-on disks. Hence, even in the worst case scenario (i.e., both failed SN candidates viewed edge-on), the remnants would still be significantly fainter than their progenitors and nothing like the much more luminous stellar merger remnants.

The absence of measurements at wavelengths longer than 21~\micron\ for the two failed SN candidates could lead to an underestimate of the remnant bolometric luminosities. However, it is very unlikely that the unobserved emission at redder wavelengths can contribute enough to increase the remnant luminosities even to the progenitor luminosities. The main argument is the necessary mass and energy budgets. To have substantial emission at longer wavelengths, the dust must be more distant. Ignoring Plank factors, the dust radius scales with luminosity $L$ and temperature $T$ as $r \propto L^{1/2}/T^2$ so the required mass scales as $M \propto r^2 \propto L/T^4$ and the kinetic energy scales as $E \propto M v^2 \propto r^4 \propto L^2/T^8$ since the required velocity $v\propto r$. In order to have 100 times more luminosity at half the temperature, the required mass would be 1600 times larger and the required kinetic energy increases a factor $2.56 \times 10^6$. This is also an underestimation of the problem because the more distant cooler dust is (presumably) being illuminated by the emission from the optically thick hotter dust, but mid-IR dust opacities are tiny compared to UV/optical opacities, so the mass also has to be higher than estimated above in order to compensate for the drop in the opacity. Nonetheless, it will be useful to observe both remnants at longer wavelengths to better characterize their SEDs (MIRI's longest wavelength F2550W filter could constrain the emission to $\sim$28~\micron). 

Another source of uncertainty can be the progenitor luminosity estimate. Apart from limited photometry in archival surveys, some stellar mergers have evidence for progenitor optical brightening years before the merger (for example, V1309~Sco \citealt{Tylenda2011} or AT~2015dl \citealt{Blagorodnova2017}). If this were true for some of the merger progenitors, then we would have underestimated the true $L_\text{prog}/L_{remn}$ ratio and so the difference from the failed SN candidates would be even larger than shown in Figure~\ref{fig:progremn}. However, most of the mergers with long archival coverage have stable luminosities over decades (see, e.g., BLG-360, \citealt{Tylenda2013}; V838~Mon, \citealt{Goranskij2004}; AT~2021biy, \citealt{Cai2022III}; or AT~2021blu, \citealt{Pastorello2023IV}). Regarding the failed SN candidates, the archival photometric coverage for \bh1 and M31-2014-DS1 prior to disappearance do not show significant long-term variability. Hence, this potential bias may be less of a concern for the failed SN candidates.

While the extragalactic stellar merger luminosities are not estimated using our own \texttt{DUSTY} spherical models, their remnant to progenitor luminosity ratios are similar to the Galactic examples and also $\sim$100 times larger than the failed SN candidates luminosity ratios. AT~2020hat stands out as the stellar merger with the largest $L_\text{remn}/L_\text{prog}$ ratio, but it is also the system with the youngest remnant luminosity measurement. Since it was measured only 100 days after the outburst, it is likely still in the plateau stage before the luminosity drops \citep[see, e.g.,][for typical LRN luminosity evolution]{Pastorello2023IV}.

The extragalactic mergers are generally from more luminous progenitors, but this is simply a selection effect. Mergers of lower mass stars are much more common than those of higher mass stars \citep[see][]{Kochanek2014}, so the local Galactic sample is dominated by lower mass, lower luminosity systems which have an appreciable rate in a single galaxy, while the extragalactic sample is dominated by higher mass, higher luminosity systems that can be detected at extragalactic distances in transient surveys of many galaxies.

Neither of the two failed SN candidates exhibit clearly detectable emission at the near-infrared (NIR) or optical wavelengths. The \texttt{DUSTY} models predict ratios between the $K_s$ band ($\lambda_\text{eff}=2.2$ \micron) and the W4 band ($\lambda_\text{eff}=22$ \micron) luminosities of $\sim$0.001. The ratio between the luminosities in the I band wavelength ($\lambda_\text{eff}=0.8$ \micron) and W4 drops basically to zero. This is not what is commonly observed for LRNe. Seven of the eight systems with wavelength coverage in these bands show NIR emission (V838~Mon, V1309~Sco, V4332~Sgr, AT~2015dl, AT~2019zhd, AT~2020hat, and AT~2021blu). The exception is BLG-360 whose SED dramatically falls bluewards of $\sim$2 \micron. For the Galactic stellar mergers, \texttt{DUSTY} models give a 2.2 \micron\ to 22 \micron\ luminosity ratio ranging from 0.008 to 1.8. Among the LRNe with NIR emission, five of them (V838~Mon, V1309~Sco, V4332~Sgr, AT~2019zhd, and AT~2020hat) also have a considerable emission in the optical with 0.8 \micron\ to 22 \micron\ luminosity ratios for the Galactic stellar mergers from 0.005 to 0.5. The SED of V838~Mon is one of the most peculiar given that its emission peaks at $\sim$2 \micron\ rather than at $\sim$10-20 \micron\ as in the other Galactic stellar mergers.

\section{Conclusions}\label{sec:concl}

We have reanalyzed the JWST photometry from Program
2896 (PI: Kochanek) and Program 3773 (PI: Beasor) for the failed SN candidate \bh1. Using the image subtraction package \texttt{ISIS}, we locate the source position accurately and find that \cite{Beasor2024} misidentified the source position in the NIRCam images (see Figure~\ref{fig:source}). Using \texttt{DOLPHOT} PSF photometry extraction procedures for the NIRCam images, we find four stars bracketing the source position in the shorter wavelength NIRCam filters (F115W, F182M, and F250M). One of these neighbors (labeled as N1) was misidentified as NIR emission from \bh1 by \cite{Beasor2024}. By fitting model stellar atmospheres with \texttt{DUSTY} inside a MCMC driver, we can characterize the four neighbors as red giant stars in NGC~6946 with luminosities ranging from $10^{2.6}$ to $10^{3.5}L_\odot$ and temperatures from 3500 to 6400~K (see Table~\ref{tab:neigparam}).

The remaining bands (NIRCam F360M and MIRI F560W, F770W, F1000W, and F2100W) are dominated by \bh1. We fit the \texttt{DOLPHOT} fluxes (see Table~\ref{tab:bh1mags}) along with upper limits from the N1 neighbor in NIRCam bands and HST measurements from \cite{Kochanek2024} with \texttt{DUSTY} models and different dust compositions and grain sizes. The best-fit model is a silicate dust shell with 668~K at its inner radius of $10^{15.42}$~cm, visual optical depth of $\tau_V$=19.4, a maximum grain size of 3~\micron\ and a total luminosity of $10^{4.73}L_\odot$. The estimated order of magnitude of the ejecta velocity is 60~km/s, with lower limits for the total ejected mass of 0.08$M_\odot$ and kinetic energy of $3\times10^{-6}$~FOE. This estimates are well below the SN explosion expectations and in line with failed SN predictions. As explained in \cite{Basinger2021}, their non-detection of X-ray emission is still consistent with BH accretion being the underlying source of the luminosity because these ejecta are Compton thick at these phases.

The models disfavor graphitic dusts as the material surrounding the object and favor a silicate dust composition due to the steep drop of the emission at $\sim$2~\micron\ and the absorption feature at $\sim$10~\micron\ (see Figure~\ref{fig:dustmod}). \cite{De2026} present a MIRI low resolution spectrum covering this spectral range for M31-2014-DS1 and found that the shape is also consistent with silicate absorption in the 8-13~\micron\ range and with no evidence for PAH emission features. Since \bh1 and M31-2014-DS1 are nearly twins, the MIR SED features of \bh1 are likely associated with silicate absorption rather than PAH emission at 7.7~\micron\ as \cite{Beasor2024} argue.

We update the $R$-band light curve of \bh1 with the ongoing LBT monitoring campaign (see Figure~\ref{fig:LBTlc}). The seven new measurements from October 2023 to September 2025 are still compatible with no detectable emission in the optical. The light curve standard deviation after the outburst is smaller than $10^3L_\odot$ without increasing or decreasing trends over these 15 years.

We compile the available examples of stellar mergers (Galactic and extragalactic) to compare their luminosities before and after the merger with the two failed SN candidates, \bh1 and M31-2014-DS1. We model the progenitors and remnants SEDs of the four Galactic stellar mergers with \texttt{DUSTY} (see Table~\ref{tab:summary} and Figure~\ref{fig:mergers}) while the luminosities for the seven extragalactic objects are taken from the literature (see Table~\ref{tab:extramergers}).

The most distinctive difference between failed SN candidates and stellar mergers is the luminosity ratio between the remnant and the progenitor. Failed SN candidates have remnants $\sim$10 times fainter than their progenitors years after the event, while stellar mergers have remnants that are $\sim$10-100 times more luminous even when the elapsed time since the outburst is similar or longer, as shown in Figure~\ref{fig:progremn}. The spherically symmetric dust distribution used to model the SEDs can underestimate the inferred luminosity by up to a factor of $\sim$3 in extreme cases where the dust is a very optically thick disk structure viewed almost exactly edge-on \citep{Kochanek2024dusty}. Even if this is the case for both failed SN candidates, this factor 3 cannot explain the roughly two orders of magnitude that separate the remnant to progenitor luminosity ratios of the failed SN candidates and the stellar mergers.

Additionally, seven out of eight stellar mergers have detectable NIR emission and five of them show optical emission as well. The V838~Mon remnant is the most remarkable case, with its  emission peaking around 2 \micron\ rather than at longer wavelengths. On the other hand, the emission of the failed SN candidates rapidly decreases for wavelengths shorter than 2 \micron. Whatever their origin, the failed SN candidates are clearly a different class of objects than stellar mergers.

\acknowledgements{
We thank Josh Peltonen, Erik Rosolowsky, and Sumit Sarbadhicary for their help with \texttt{DOLPHOT}. CSK and RFT are supported by NSF grants AST-2307385 and 2407206. The LBT is an international collaboration among institutions in the United States, Italy, and Germany. LBT Corporation partners are: The University of Arizona on behalf of the Arizona Board of Regents; Istituto Nazionale di Astrofisica, Italy; LBT Beteiligungsgesellschaft, Germany, representing the Max-Planck Society, The Leibniz Institute for Astrophysics Potsdam, and Heidelberg University; The Ohio State University, representing OSU, University of Notre Dame, University of Minnesota, and University of Virginia.
}

\bibliographystyle{aasjournal}
\bibliography{main}{}

@ARTICLE{Beasor2024,
       author = {{Beasor}, Emma R. and {Hosseinzadeh}, Griffin and {Smith}, Nathan and {Davies}, Ben and {Jencson}, Jacob E. and {Pearson}, Jeniveve and {Sand}, David J.},
        title = "{JWST Reveals a Luminous Infrared Source at the Position of the Failed Supernova Candidate N6946-BH1}",
      journal = {\apj},
         year = 2024,
        month = apr,
       volume = {964},
       number = {2},
          eid = {171},
        pages = {171},
          doi = {10.3847/1538-4357/ad21fa},
archivePrefix = {arXiv},
       eprint = {2309.16121},
 primaryClass = {astro-ph.SR},
       adsurl = {https://ui.adsabs.harvard.edu/abs/2024ApJ...964..171B}
}

@ARTICLE{Kochanek2024,
       author = {{Kochanek}, Christopher S. and {Neustadt}, Jack M.~M. and {Stanek}, Krzysztof Z.},
        title = "{The Search for Failed Supernovae with the Large Binocular Telescope: The Mid-infrared Counterpart to N6946-BH1}",
      journal = {\apj},
         year = 2024,
        month = feb,
       volume = {962},
       number = {2},
          eid = {145},
        pages = {145},
          doi = {10.3847/1538-4357/ad18d7},
archivePrefix = {arXiv},
       eprint = {2310.01514},
 primaryClass = {astro-ph.HE},
       adsurl = {https://ui.adsabs.harvard.edu/abs/2024ApJ...962..145K}
}

@ARTICLE{Dolphin2000,
       author = {{Dolphin}, Andrew E.},
        title = "{WFPC2 Stellar Photometry with HSTPHOT}",
      journal = {\pasp},
         year = 2000,
        month = oct,
       volume = {112},
       number = {776},
        pages = {1383-1396},
          doi = {10.1086/316630},
archivePrefix = {arXiv},
       eprint = {astro-ph/0006217},
 primaryClass = {astro-ph},
       adsurl = {https://ui.adsabs.harvard.edu/abs/2000PASP..112.1383D}
}

@software{Dolphin2016,
       author = {{Dolphin}, Andrew},
        title = "{DOLPHOT: Stellar photometry}",
 howpublished = {Astrophysics Source Code Library, record ascl:1608.013},
         year = 2016,
        month = aug,
          eid = {ascl:1608.013},
archivePrefix = {ascl},
       eprint = {1608.013},
       adsurl = {https://ui.adsabs.harvard.edu/abs/2016ascl.soft08013D}
}

@ARTICLE{Weisz2024,
       author = {{Weisz}, Daniel R. and {Dolphin}, Andrew E. and {Savino}, Alessandro and {McQuinn}, Kristen B.~W. and {Newman}, Max J.~B. and {Williams}, Benjamin F. and {Kallivayalil}, Nitya and {Anderson}, Jay and {Boyer}, Martha L. and {Correnti}, Matteo and {Geha}, Marla C. and {Sandstrom}, Karin M. and {Cole}, Andrew A. and {Warfield}, Jack T. and {Skillman}, Evan D. and {Cohen}, Roger E. and {Beaton}, Rachael and {Bressan}, Alessandro and {Bolatto}, Alberto and {Boylan-Kolchin}, Michael and {Brooks}, Alyson M. and {Bullock}, James S. and {Conroy}, Charlie and {Cooper}, Michael C. and {Dalcanton}, Julianne J. and {Dotter}, Aaron L. and {Fritz}, Tobias K. and {Garling}, Christopher T. and {Gennaro}, Mario and {Gilbert}, Karoline M. and {Girardi}, Leo and {Johnson}, Benjamin D. and {Johnson}, L. Clifton and {Kalirai}, Jason and {Kirby}, Evan N. and {Lang}, Dustin and {Marigo}, Paola and {Richstein}, Hannah and {Schlafly}, Edward F. and {Tollerud}, Erik J. and {Wetzel}, Andrew},
        title = "{The JWST Resolved Stellar Populations Early Release Science Program. V. DOLPHOT Stellar Photometry for NIRCam and NIRISS}",
      journal = {\apjs},
         year = 2024,
        month = apr,
       volume = {271},
       number = {2},
          eid = {47},
        pages = {47},
          doi = {10.3847/1538-4365/ad2600},
archivePrefix = {arXiv},
       eprint = {2402.03504},
 primaryClass = {astro-ph.GA},
       adsurl = {https://ui.adsabs.harvard.edu/abs/2024ApJS..271...47W}
}

@ARTICLE{Alard1998,
       author = {{Alard}, C. and {Lupton}, Robert H.},
        title = "{A Method for Optimal Image Subtraction}",
      journal = {\apj},
         year = 1998,
        month = aug,
       volume = {503},
       number = {1},
        pages = {325-331},
          doi = {10.1086/305984},
archivePrefix = {arXiv},
       eprint = {astro-ph/9712287},
 primaryClass = {astro-ph},
       adsurl = {https://ui.adsabs.harvard.edu/abs/1998ApJ...503..325A}
}

@ARTICLE{Alard2000,
       author = {{Alard}, C.},
        title = "{Image subtraction using a space-varying kernel}",
      journal = {\aaps},
         year = 2000,
        month = jun,
       volume = {144},
        pages = {363-370},
          doi = {10.1051/aas:2000214},
       adsurl = {https://ui.adsabs.harvard.edu/abs/2000A&AS..144..363A}
}

@ARTICLE{Basinger2021,
       author = {{Basinger}, C.~M. and {Kochanek}, C.~S. and {Adams}, S.~M. and {Dai}, X. and {Stanek}, K.~Z.},
        title = "{The search for failed supernovae with the Large Binocular Telescope: N6946-BH1, still no star}",
      journal = {\mnras},
         year = 2021,
        month = nov,
       volume = {508},
       number = {1},
        pages = {1156-1164},
          doi = {10.1093/mnras/stab2620},
archivePrefix = {arXiv},
       eprint = {2007.15658},
 primaryClass = {astro-ph.SR},
       adsurl = {https://ui.adsabs.harvard.edu/abs/2021MNRAS.508.1156B}
}

@ARTICLE{Schlafly2011,
       author = {{Schlafly}, Edward F. and {Finkbeiner}, Douglas P.},
        title = "{Measuring Reddening with Sloan Digital Sky Survey Stellar Spectra and Recalibrating SFD}",
      journal = {\apj},
         year = 2011,
        month = aug,
       volume = {737},
       number = {2},
          eid = {103},
        pages = {103},
          doi = {10.1088/0004-637X/737/2/103},
archivePrefix = {arXiv},
       eprint = {1012.4804},
 primaryClass = {astro-ph.GA},
       adsurl = {https://ui.adsabs.harvard.edu/abs/2011ApJ...737..103S}
}

@ARTICLE{Anand2018,
       author = {{Anand}, Gagandeep S. and {Rizzi}, Luca and {Tully}, R. Brent},
        title = "{A Robust Tip of the Red Giant Branch Distance to the Fireworks Galaxy (NGC 6946)}",
      journal = {\aj},
         year = 2018,
        month = sep,
       volume = {156},
       number = {3},
          eid = {105},
        pages = {105},
          doi = {10.3847/1538-3881/aad3b2},
archivePrefix = {arXiv},
       eprint = {1807.05229},
 primaryClass = {astro-ph.GA},
       adsurl = {https://ui.adsabs.harvard.edu/abs/2018AJ....156..105A}
}

@ARTICLE{Gustafsson2008,
       author = {{Gustafsson}, B. and {Edvardsson}, B. and {Eriksson}, K. and {J{\o}rgensen}, U.~G. and {Nordlund}, {\r{A}}. and {Plez}, B.},
        title = "{A grid of MARCS model atmospheres for late-type stars. I. Methods and general properties}",
      journal = {\aap},
         year = 2008,
        month = aug,
       volume = {486},
       number = {3},
        pages = {951-970},
          doi = {10.1051/0004-6361:200809724},
archivePrefix = {arXiv},
       eprint = {0805.0554},
 primaryClass = {astro-ph},
       adsurl = {https://ui.adsabs.harvard.edu/abs/2008A&A...486..951G}
}

@INPROCEEDINGS{Castelli2003,
       author = {{Castelli}, F. and {Kurucz}, R.~L.},
        title = "{New Grids of ATLAS9 Model Atmospheres}",
    booktitle = {Modelling of Stellar Atmospheres},
         year = 2003,
       editor = {{Piskunov}, N. and {Weiss}, W.~W. and {Gray}, D.~F.},
       series = {IAU Symposium},
       volume = {210},
        month = jan,
        pages = {A20},
          doi = {10.48550/arXiv.astro-ph/0405087},
archivePrefix = {arXiv},
       eprint = {astro-ph/0405087},
 primaryClass = {astro-ph},
       adsurl = {https://ui.adsabs.harvard.edu/abs/2003IAUS..210P.A20C}
}

@ARTICLE{Elitzur2001,
       author = {{Elitzur}, Moshe and {Ivezi{\'c}}, {\v{Z}}eljko},
        title = "{Dusty winds - I. Self-similar solutions}",
      journal = {\mnras},
         year = 2001,
        month = oct,
       volume = {327},
       number = {2},
        pages = {403-421},
          doi = {10.1046/j.1365-8711.2001.04706.x},
archivePrefix = {arXiv},
       eprint = {astro-ph/0106096},
 primaryClass = {astro-ph},
       adsurl = {https://ui.adsabs.harvard.edu/abs/2001MNRAS.327..403E}
}

@ARTICLE{Draine1984,
       author = {{Draine}, B.~T. and {Lee}, H.~M.},
        title = "{Optical Properties of Interstellar Graphite and Silicate Grains}",
      journal = {\apj},
         year = 1984,
        month = oct,
       volume = {285},
        pages = {89},
          doi = {10.1086/162480},
       adsurl = {https://ui.adsabs.harvard.edu/abs/1984ApJ...285...89D}
}

@ARTICLE{Mathis1977,
       author = {{Mathis}, J.~S. and {Rumpl}, W. and {Nordsieck}, K.~H.},
        title = "{The size distribution of interstellar grains.}",
      journal = {\apj},
         year = 1977,
        month = oct,
       volume = {217},
        pages = {425-433},
          doi = {10.1086/155591},
       adsurl = {https://ui.adsabs.harvard.edu/abs/1977ApJ...217..425M}
}

@ARTICLE{Adams2017BH1,
       author = {{Adams}, S.~M. and {Kochanek}, C.~S. and {Gerke}, J.~R. and {Stanek}, K.~Z. and {Dai}, X.},
        title = "{The search for failed supernovae with the Large Binocular Telescope: confirmation of a disappearing star}",
      journal = {\mnras},
         year = {2017b},
        month = jul,
       volume = {468},
       number = {4},
        pages = {4968-4981},
          doi = {10.1093/mnras/stx816},
archivePrefix = {arXiv},
       eprint = {1609.01283},
 primaryClass = {astro-ph.SR},
       adsurl = {https://ui.adsabs.harvard.edu/abs/2017MNRAS.468.4968A}
}

@ARTICLE{Beasor2026,
       author = {{Beasor}, Emma R. and {Smith}, Nathan and {Pearson}, Jeniveve and {Subrayan}, Bhagya and {Berger}, Edo and {Sand}, David J. and {Strader}, Jay},
        title = "{The fate of the failed supernova candidate M31-2014-DS1}",
      journal = {\mnras},
         year = 2026,
        month = jan,
          doi = {10.1093/mnras/stag052},
archivePrefix = {arXiv},
       eprint = {2601.05317},
 primaryClass = {astro-ph.SR},
       adsurl = {https://ui.adsabs.harvard.edu/abs/2026MNRAS.tmp...60B}
}

@ARTICLE{Shara1985,
       author = {{Shara}, M.~M. and {Moffat}, A.~F.~J. and {Webbink}, R.~F.},
        title = "{Unraveling the oldest and faintest recovered nova : CK Vulpeculae (1670).}",
      journal = {\apj},
         year = 1985,
        month = jul,
       volume = {294},
        pages = {271-285},
          doi = {10.1086/163296},
       adsurl = {https://ui.adsabs.harvard.edu/abs/1985ApJ...294..271S}
}

@ARTICLE{Kaminski2015,
       author = {{Kami{\'n}ski}, Tomasz and {Menten}, Karl M. and {Tylenda}, Romuald and {Hajduk}, Marcin and {Patel}, Nimesh A. and {Kraus}, Alexander},
        title = "{Nuclear ashes and outflow in the eruptive star Nova Vul 1670}",
      journal = {\nat},
         year = 2015,
        month = apr,
       volume = {520},
       number = {7547},
        pages = {322-324},
          doi = {10.1038/nature14257},
archivePrefix = {arXiv},
       eprint = {1503.06570},
 primaryClass = {astro-ph.SR},
       adsurl = {https://ui.adsabs.harvard.edu/abs/2015Natur.520..322K}
}

@ARTICLE{Kaminski2021,
       author = {{Kami{\'n}ski}, T. and {Steffen}, W. and {Bujarrabal}, V. and {Tylenda}, R. and {Menten}, K.~M. and {Hajduk}, M.},
        title = "{Molecular remnant of Nova 1670 (CK Vulpeculae). II. A three-dimensional view of the gas distribution and velocity field}",
      journal = {\aap},
         year = 2021,
        month = feb,
       volume = {646},
          eid = {A1},
        pages = {A1},
          doi = {10.1051/0004-6361/202039634},
archivePrefix = {arXiv},
       eprint = {2010.05832},
 primaryClass = {astro-ph.SR},
       adsurl = {https://ui.adsabs.harvard.edu/abs/2021A&A...646A...1K}
}

@ARTICLE{Tylenda2024,
       author = {{Tylenda}, Romuald and {Kami{\'n}ski}, Tomek and {Smolec}, Radek},
        title = "{Nova 1670 (CK Vulpeculae) was a merger of a red giant with a helium white dwarf}",
      journal = {\aap},
         year = 2024,
        month = may,
       volume = {685},
          eid = {A49},
        pages = {A49},
          doi = {10.1051/0004-6361/202244896},
archivePrefix = {arXiv},
       eprint = {2312.07433},
 primaryClass = {astro-ph.SR},
       adsurl = {https://ui.adsabs.harvard.edu/abs/2024A&A...685A..49T}
}

@ARTICLE{Banerjee2020,
       author = {{Banerjee}, D.~P.~K. and {Geballe}, T.~R. and {Evans}, A. and {Shahbandeh}, M. and {Woodward}, C.~E. and {Gehrz}, R.~D. and {Eyres}, S.~P.~S. and {Starrfield}, S. and {Zijlstra}, A.},
        title = "{Near-infrared Spectroscopy of CK Vulpeculae: Revealing a Remarkably Powerful Blast from the Past}",
      journal = {\apjl},
         year = 2020,
        month = dec,
       volume = {904},
       number = {2},
          eid = {L23},
        pages = {L23},
          doi = {10.3847/2041-8213/abc885},
archivePrefix = {arXiv},
       eprint = {2011.02939},
 primaryClass = {astro-ph.SR},
       adsurl = {https://ui.adsabs.harvard.edu/abs/2020ApJ...904L..23B}
}

@ARTICLE{Kato2003,
       author = {{Kato}, T.},
        title = "{CK Vul as a candidate eruptive stellar merging event}",
      journal = {\aap},
         year = 2003,
        month = feb,
       volume = {399},
        pages = {695-697},
          doi = {10.1051/0004-6361:20021808},
archivePrefix = {arXiv},
       eprint = {astro-ph/0211557},
 primaryClass = {astro-ph},
       adsurl = {https://ui.adsabs.harvard.edu/abs/2003A&A...399..695K}
}

@ARTICLE{Hajduk2013,
       author = {{Hajduk}, M. and {van Hoof}, P.~A.~M. and {Zijlstra}, A.~A.},
        title = "{CK Vul: evolving nebula and three curious background stars}",
      journal = {\mnras},
         year = 2013,
        month = jun,
       volume = {432},
       number = {1},
        pages = {167-175},
          doi = {10.1093/mnras/stt426},
archivePrefix = {arXiv},
       eprint = {1312.5846},
 primaryClass = {astro-ph.SR},
       adsurl = {https://ui.adsabs.harvard.edu/abs/2013MNRAS.432..167H}
}

@ARTICLE{Evans2016,
       author = {{Evans}, A. and {Gehrz}, R.~D. and {Woodward}, C.~E. and {Sarre}, P.~J. and {van Loon}, J.~T. and {Helton}, L.~A. and {Starrfield}, S. and {Eyres}, S.~P.~S.},
        title = "{CK Vul: a smorgasbord of hydrocarbons rules out a 1670 nova (and much else besides)}",
      journal = {\mnras},
         year = 2016,
        month = apr,
       volume = {457},
       number = {3},
        pages = {2871-2876},
          doi = {10.1093/mnras/stw352},
archivePrefix = {arXiv},
       eprint = {1512.02146},
 primaryClass = {astro-ph.SR},
       adsurl = {https://ui.adsabs.harvard.edu/abs/2016MNRAS.457.2871E}
}

@ARTICLE{Eyres2018,
       author = {{Eyres}, S.~P.~S. and {Evans}, A. and {Zijlstra}, A. and {Avison}, A. and {Gehrz}, R.~D. and {Hajduk}, M. and {Starrfield}, S. and {Mohamed}, S. and {Woodward}, C.~E. and {Wagner}, R.~M.},
        title = "{ALMA reveals the aftermath of a white dwarf-brown dwarf merger in CK Vulpeculae}",
      journal = {\mnras},
         year = 2018,
        month = dec,
       volume = {481},
       number = {4},
        pages = {4931-4939},
          doi = {10.1093/mnras/sty2554},
archivePrefix = {arXiv},
       eprint = {1809.05849},
 primaryClass = {astro-ph.SR},
       adsurl = {https://ui.adsabs.harvard.edu/abs/2018MNRAS.481.4931E}
}

@ARTICLE{Tylenda2013,
       author = {{Tylenda}, R. and {Kami{\'n}ski}, T. and {Udalski}, A. and {Soszy{\'n}ski}, I. and {Poleski}, R. and {Szyma{\'n}ski}, M.~K. and {Kubiak}, M. and {Pietrzy{\'n}ski}, G. and {Koz{\l}owski}, S. and {Pietrukowicz}, P. and {Ulaczyk}, K. and {Wyrzykowski}, {\L}.},
        title = "{OGLE-2002-BLG-360: from a gravitational microlensing candidate to an overlooked red transient}",
      journal = {\aap},
         year = 2013,
        month = jul,
       volume = {555},
          eid = {A16},
        pages = {A16},
          doi = {10.1051/0004-6361/201321647},
archivePrefix = {arXiv},
       eprint = {1304.1694},
 primaryClass = {astro-ph.SR},
       adsurl = {https://ui.adsabs.harvard.edu/abs/2013A&A...555A..16T}
}

@ARTICLE{Steinmetz2025,
       author = {{Steinmetz}, T. and {Kami{\'n}ski}, T. and {Melis}, C. and {Blagorodnova}, N. and {Gromadzki}, M. and {Menten}, K. and {Su}, K.},
        title = "{OGLE-2002-BLG-360: A dusty anomaly among red nova remnants}",
      journal = {\aap},
         year = 2025,
        month = jul,
       volume = {699},
          eid = {A316},
        pages = {A316},
          doi = {10.1051/0004-6361/202554261},
archivePrefix = {arXiv},
       eprint = {2502.18365},
 primaryClass = {astro-ph.SR},
       adsurl = {https://ui.adsabs.harvard.edu/abs/2025A&A...699A.316S}
}

@ARTICLE{Brown2002,
       author = {{Brown}, N.~J. and {Waagen}, E.~O. and {Scovil}, C. and {Nelson}, P. and {Oksanen}, A. and {Solonen}, J. and {Price}, A.},
        title = "{Peculiar variable in Monoceros.}",
      journal = {\iaucirc},
         year = 2002,
        month = jan,
       volume = {7785},
        pages = {1},
       adsurl = {https://ui.adsabs.harvard.edu/abs/2002IAUC.7785....1B}
}

@ARTICLE{Woodward2021,
       author = {{Woodward}, C.~E. and {Evans}, A. and {Banerjee}, D.~P.~K. and {Liimets}, T. and {Djupvik}, A.~A. and {Starrfield}, S. and {Clayton}, G.~C. and {Eyres}, S.~P.~S. and {Gehrz}, R.~D. and {Wagner}, R.~M.},
        title = "{The Infrared Evolution of Dust in V838 Monocerotis}",
      journal = {\aj},
         year = 2021,
        month = nov,
       volume = {162},
       number = {5},
          eid = {183},
        pages = {183},
          doi = {10.3847/1538-3881/ac1f1e},
archivePrefix = {arXiv},
       eprint = {2108.08149},
 primaryClass = {astro-ph.SR},
       adsurl = {https://ui.adsabs.harvard.edu/abs/2021AJ....162..183W}
}

@ARTICLE{Tylenda2005V838Mon,
       author = {{Tylenda}, R. and {Soker}, N. and {Szczerba}, R.},
        title = "{On the progenitor of V838 Monocerotis}",
      journal = {\aap},
         year = 2005,
        month = oct,
       volume = {441},
       number = {3},
        pages = {1099-1109},
          doi = {10.1051/0004-6361:20042485},
archivePrefix = {arXiv},
       eprint = {astro-ph/0412183},
 primaryClass = {astro-ph},
       adsurl = {https://ui.adsabs.harvard.edu/abs/2005A&A...441.1099T}
}

@ARTICLE{Sparks2008,
       author = {{Sparks}, William B. and {Bond}, Howard E. and {Cracraft}, Misty and {Levay}, Zolt and {Crause}, Lisa A. and {Dopita}, Michael A. and {Henden}, Arne A. and {Munari}, Ulisse and {Panagia}, Nino and {Starrfield}, Sumner G. and {Sugerman}, Ben E. and {Wagner}, R. Mark and {White}, Richard L.},
        title = "{V838 Monocerotis: A Geometric Distance from Hubble Space Telescope Polarimetric Imaging of its Light Echo}",
      journal = {\aj},
         year = 2008,
        month = feb,
       volume = {135},
       number = {2},
        pages = {605-617},
          doi = {10.1088/0004-6256/135/2/605},
archivePrefix = {arXiv},
       eprint = {0711.1495},
 primaryClass = {astro-ph},
       adsurl = {https://ui.adsabs.harvard.edu/abs/2008AJ....135..605S}
}

@ARTICLE{Nakano2008,
       author = {{Nakano}, S. and {Nishiyama}, K. and {Kabashima}, F. and {Sakurai}, Y.},
        title = "{Possible Nova in Scorpius}",
      journal = {Central Bureau Electronic Telegrams},
         year = 2008,
        month = sep,
       volume = {1496},
        pages = {1},
       adsurl = {https://ui.adsabs.harvard.edu/abs/2008CBET.1496....1N}
}

@ARTICLE{Mason2010,
       author = {{Mason}, E. and {Diaz}, M. and {Williams}, R.~E. and {Preston}, G. and {Bensby}, T.},
        title = "{The peculiar nova V1309 Scorpii/nova Scorpii 2008. A candidate twin of V838 Monocerotis}",
      journal = {\aap},
         year = 2010,
        month = jun,
       volume = {516},
          eid = {A108},
        pages = {A108},
          doi = {10.1051/0004-6361/200913610},
archivePrefix = {arXiv},
       eprint = {1004.3600},
 primaryClass = {astro-ph.SR},
       adsurl = {https://ui.adsabs.harvard.edu/abs/2010A&A...516A.108M}
}

@ARTICLE{Tylenda2011,
       author = {{Tylenda}, R. and {Hajduk}, M. and {Kami{\'n}ski}, T. and {Udalski}, A. and {Soszy{\'n}ski}, I. and {Szyma{\'n}ski}, M.~K. and {Kubiak}, M. and {Pietrzy{\'n}ski}, G. and {Poleski}, R. and {Wyrzykowski}, {\L}. and {Ulaczyk}, K.},
        title = "{V1309 Scorpii: merger of a contact binary}",
      journal = {\aap},
         year = 2011,
        month = apr,
       volume = {528},
          eid = {A114},
        pages = {A114},
          doi = {10.1051/0004-6361/201016221},
archivePrefix = {arXiv},
       eprint = {1012.0163},
 primaryClass = {astro-ph.SR},
       adsurl = {https://ui.adsabs.harvard.edu/abs/2011A&A...528A.114T}
}

@ARTICLE{Tylenda2016,
       author = {{Tylenda}, R. and {Kami{\'n}ski}, T.},
        title = "{Evolution of the stellar-merger red nova V1309 Scorpii: Spectral energy distribution analysis}",
      journal = {\aap},
         year = 2016,
        month = aug,
       volume = {592},
          eid = {A134},
        pages = {A134},
          doi = {10.1051/0004-6361/201527700},
archivePrefix = {arXiv},
       eprint = {1606.09426},
 primaryClass = {astro-ph.SR},
       adsurl = {https://ui.adsabs.harvard.edu/abs/2016A&A...592A.134T}
}

@ARTICLE{McCollum2014,
       author = {{McCollum}, Bruce and {Laine}, Seppo and {V{\"a}is{\"a}nen}, Petri and {Bruhweiler}, Frederick C. and {Rottler}, Lee and {Ryder}, Stuart and {Wahlgren}, Glenn M. and {Barway}, Sudhanshu and {Nagayama}, Takahiro and {Ramphul}, Rajin},
        title = "{The Optical and Infrared Photometric Evolution of the Recent Stellar Merger, V1309 Sco}",
      journal = {\aj},
         year = 2014,
        month = jan,
       volume = {147},
       number = {1},
          eid = {11},
        pages = {11},
          doi = {10.1088/0004-6256/147/1/11},
       adsurl = {https://ui.adsabs.harvard.edu/abs/2014AJ....147...11M}
}

@ARTICLE{Hayashi1994,
       author = {{Hayashi}, S.~S. and {Yamamoto}, M. and {Hirosawa}, K.},
        title = "{Nova Sagittarii 1994}",
      journal = {\iaucirc},
         year = 1994,
        month = feb,
       volume = {5942},
        pages = {1},
       adsurl = {https://ui.adsabs.harvard.edu/abs/1994IAUC.5942....1H}
}

@ARTICLE{Martini1999,
       author = {{Martini}, Paul and {Wagner}, R. Mark and {Tomaney}, Austin and {Rich}, R. Michael and {della Valle}, M. and {Hauschildt}, Peter H.},
        title = "{Nova Sagittarii 1994 1 (V4332 Sagittarii): The Discovery and Evolution of an Unusual Luminous Red Variable Star}",
      journal = {\aj},
         year = 1999,
        month = aug,
       volume = {118},
       number = {2},
        pages = {1034-1042},
          doi = {10.1086/300951},
archivePrefix = {arXiv},
       eprint = {astro-ph/9905016},
 primaryClass = {astro-ph},
       adsurl = {https://ui.adsabs.harvard.edu/abs/1999AJ....118.1034M}
}

@ARTICLE{Tylenda2005V4332Sgr,
       author = {{Tylenda}, R. and {Crause}, L.~A. and {G{\'o}rny}, S.~K. and {Schmidt}, M.~R.},
        title = "{V4332 Sagittarii revisited}",
      journal = {\aap},
         year = 2005,
        month = aug,
       volume = {439},
       number = {2},
        pages = {651-661},
          doi = {10.1051/0004-6361:20041581},
archivePrefix = {arXiv},
       eprint = {astro-ph/0412205},
 primaryClass = {astro-ph},
       adsurl = {https://ui.adsabs.harvard.edu/abs/2005A&A...439..651T}
}

@ARTICLE{Kaminski2010,
       author = {{Kami{\'n}ski}, T. and {Schmidt}, M. and {Tylenda}, R.},
        title = "{V4332 Sagittarii: a circumstellar disc obscuring the main object}",
      journal = {\aap},
         year = 2010,
        month = nov,
       volume = {522},
          eid = {A75},
        pages = {A75},
          doi = {10.1051/0004-6361/201014406},
archivePrefix = {arXiv},
       eprint = {1007.0131},
 primaryClass = {astro-ph.SR},
       adsurl = {https://ui.adsabs.harvard.edu/abs/2010A&A...522A..75K}
}

@ARTICLE{Blagorodnova2017,
       author = {{Blagorodnova}, N. and {Kotak}, R. and {Polshaw}, J. and {Kasliwal}, M.~M. and {Cao}, Y. and {Cody}, A.~M. and {Doran}, G.~B. and {Elias-Rosa}, N. and {Fraser}, M. and {Fremling}, C. and {Gonzalez-Fernandez}, C. and {Harmanen}, J. and {Jencson}, J. and {Kankare}, E. and {Kudritzki}, R.-P. and {Kulkarni}, S.~R. and {Magnier}, E. and {Manulis}, I. and {Masci}, F.~J. and {Mattila}, S. and {Nugent}, P. and {Ochner}, P. and {Pastorello}, A. and {Reynolds}, T. and {Smith}, K. and {Sollerman}, J. and {Taddia}, F. and {Terreran}, G. and {Tomasella}, L. and {Turatto}, M. and {Vreeswijk}, P.~M. and {Wozniak}, P. and {Zaggia}, S.},
        title = "{Common Envelope Ejection for a Luminous Red Nova in M101}",
      journal = {\apj},
         year = 2017,
        month = jan,
       volume = {834},
       number = {2},
          eid = {107},
        pages = {107},
          doi = {10.3847/1538-4357/834/2/107},
archivePrefix = {arXiv},
       eprint = {1607.08248},
 primaryClass = {astro-ph.SR},
       adsurl = {https://ui.adsabs.harvard.edu/abs/2017ApJ...834..107B}
}

@ARTICLE{Pastorello2023IV,
       author = {{Pastorello}, A. and {Valerin}, G. and {Fraser}, M. and {Reguitti}, A. and {Elias-Rosa}, N. and {Filippenko}, A.~V. and {Rojas-Bravo}, C. and {Tartaglia}, L. and {Reynolds}, T.~M. and {Valenti}, S. and {Andrews}, J.~E. and {Ashall}, C. and {Bostroem}, K.~A. and {Brink}, T.~G. and {Burke}, J. and {Cai}, Y.-Z. and {Cappellaro}, E. and {Coulter}, D.~A. and {Dastidar}, R. and {Davis}, K.~W. and {Dimitriadis}, G. and {Fiore}, A. and {Foley}, R.~J. and {Fugazza}, D. and {Galbany}, L. and {Gangopadhyay}, A. and {Geier}, S. and {Guti{\'e}rrez}, C.~P. and {Haislip}, J. and {Hiramatsu}, D. and {Holmbo}, S. and {Howell}, D.~A. and {Hsiao}, E.~Y. and {Hung}, T. and {Jha}, S.~W. and {Kankare}, E. and {Karamehmetoglu}, E. and {Kilpatrick}, C.~D. and {Kotak}, R. and {Kouprianov}, V. and {Kravtsov}, T. and {Kumar}, S. and {Li}, Z.-T. and {Lundquist}, M.~J. and {Lundqvist}, P. and {Matilainen}, K. and {Mazzali}, P.~A. and {McCully}, C. and {Misra}, K. and {Morales-Garoffolo}, A. and {Moran}, S. and {Morrell}, N. and {Newsome}, M. and {Padilla Gonzalez}, E. and {Pan}, Y.-C. and {Pellegrino}, C. and {Phillips}, M.~M. and {Pignata}, G. and {Piro}, A.~L. and {Reichart}, D.~E. and {Rest}, A. and {Salmaso}, I. and {Sand}, D.~J. and {Siebert}, M.~R. and {Smartt}, S.~J. and {Smith}, K.~W. and {Srivastav}, S. and {Stritzinger}, M.~D. and {Taggart}, K. and {Tinyanont}, S. and {Yan}, S.-Y. and {Wang}, L. and {Wang}, X.-F. and {Williams}, S.~C. and {Wyatt}, S. and {Zhang}, T.-M. and {de Boer}, T. and {Chambers}, K. and {Gao}, H. and {Magnier}, E.},
        title = "{Forbidden hugs in pandemic times. IV. Panchromatic evolution of three luminous red novae}",
      journal = {\aap},
         year = 2023,
        month = mar,
       volume = {671},
          eid = {A158},
        pages = {A158},
          doi = {10.1051/0004-6361/202244684},
archivePrefix = {arXiv},
       eprint = {2208.02782},
 primaryClass = {astro-ph.SR},
       adsurl = {https://ui.adsabs.harvard.edu/abs/2023A&A...671A.158P}
}

@ARTICLE{Cai2022III,
       author = {{Cai}, Y.-Z. and {Pastorello}, A. and {Fraser}, M. and {Wang}, X.-F. and {Filippenko}, A.~V. and {Reguitti}, A. and {Patra}, K.~C. and {Goranskij}, V.~P. and {Barsukova}, E.~A. and {Brink}, T.~G. and {Elias-Rosa}, N. and {Stevance}, H.~F. and {Zheng}, W. and {Yang}, Y. and {Atapin}, K.~E. and {Benetti}, S. and {de Boer}, T.~J.~L. and {Bose}, S. and {Burke}, J. and {Byrne}, R. and {Cappellaro}, E. and {Chambers}, K.~C. and {Chen}, W.-L. and {Emami}, N. and {Gao}, H. and {Hiramatsu}, D. and {Howell}, D.~A. and {Huber}, M.~E. and {Kankare}, E. and {Kelly}, P.~L. and {Kotak}, R. and {Kravtsov}, T. and {Lander}, V. Yu. and {Li}, Z.-T. and {Lin}, C.-C. and {Lundqvist}, P. and {Magnier}, E.~A. and {Malygin}, E.~A. and {Maslennikova}, N.~A. and {Matilainen}, K. and {Mazzali}, P.~A. and {McCully}, C. and {Mo}, J. and {Moran}, S. and {Newsome}, M. and {Oparin}, D.~V. and {Padilla Gonzalez}, E. and {Reynolds}, T.~M. and {Shatsky}, N.~I. and {Smartt}, S.~J. and {Smith}, K.~W. and {Stritzinger}, M.~D. and {Tatarnikov}, A.~M. and {Terreran}, G. and {Uklein}, R.~I. and {Valerin}, G. and {Vallely}, P.~J. and {Vozyakova}, O.~V. and {Wainscoat}, R. and {Yan}, S.-Y. and {Zhang}, J.-J. and {Zhang}, T.-M. and {Zheltoukhov}, S.~G. and {Dastidar}, R. and {Fulton}, M. and {Galbany}, L. and {Gangopadhyay}, A. and {Ge}, H.-W. and {Guti{\'e}rrez}, C.~P. and {Lin}, H. and {Misra}, K. and {Ou}, Z.-W. and {Salmaso}, I. and {Tartaglia}, L. and {Xiao}, L. and {Zhang}, X.-H.},
        title = "{Forbidden hugs in pandemic times. III. Observations of the luminous red nova AT 2021biy in the nearby galaxy NGC 4631}",
      journal = {\aap},
         year = 2022,
        month = nov,
       volume = {667},
          eid = {A4},
        pages = {A4},
          doi = {10.1051/0004-6361/202244393},
archivePrefix = {arXiv},
       eprint = {2207.00734},
 primaryClass = {astro-ph.SR},
       adsurl = {https://ui.adsabs.harvard.edu/abs/2022A&A...667A...4C}
}

@ARTICLE{Pastorello2021II,
       author = {{Pastorello}, A. and {Valerin}, G. and {Fraser}, M. and {Elias-Rosa}, N. and {Valenti}, S. and {Reguitti}, A. and {Mazzali}, P.~A. and {Amaro}, R.~C. and {Andrews}, J.~E. and {Dong}, Y. and {Jencson}, J. and {Lundquist}, M. and {Reichart}, D.~E. and {Sand}, D.~J. and {Wyatt}, S. and {Smartt}, S.~J. and {Smith}, K.~W. and {Srivastav}, S. and {Cai}, Y.-Z. and {Cappellaro}, E. and {Holmbo}, S. and {Fiore}, A. and {Jones}, D. and {Kankare}, E. and {Karamehmetoglu}, E. and {Lundqvist}, P. and {Morales-Garoffolo}, A. and {Reynolds}, T.~M. and {Stritzinger}, M.~D. and {Williams}, S.~C. and {Chambers}, K.~C. and {de Boer}, T.~J.~L. and {Huber}, M.~E. and {Rest}, A. and {Wainscoat}, R.},
        title = "{Forbidden hugs in pandemic times. II. The luminous red nova variety: AT 2020hat and AT 2020kog}",
      journal = {\aap},
         year = 2021,
        month = mar,
       volume = {647},
          eid = {A93},
        pages = {A93},
          doi = {10.1051/0004-6361/202039953},
archivePrefix = {arXiv},
       eprint = {2011.10590},
 primaryClass = {astro-ph.SR},
       adsurl = {https://ui.adsabs.harvard.edu/abs/2021A&A...647A..93P}
}

@ARTICLE{Pastorello2021I,
       author = {{Pastorello}, A. and {Fraser}, M. and {Valerin}, G. and {Reguitti}, A. and {Itagaki}, K. and {Ochner}, P. and {Williams}, S.~C. and {Jones}, D. and {Munday}, J. and {Smartt}, S.~J. and {Smith}, K.~W. and {Srivastav}, S. and {Elias-Rosa}, N. and {Kankare}, E. and {Karamehmetoglu}, E. and {Lundqvist}, P. and {Mazzali}, P.~A. and {Munari}, U. and {Stritzinger}, M.~D. and {Tomasella}, L. and {Anderson}, J.~P. and {Chambers}, K.~C. and {Rest}, A.},
        title = "{Forbidden hugs in pandemic times. I. Luminous red nova AT 2019zhd, a new merger in M 31}",
      journal = {\aap},
         year = 2021,
        month = feb,
       volume = {646},
          eid = {A119},
        pages = {A119},
          doi = {10.1051/0004-6361/202039952},
archivePrefix = {arXiv},
       eprint = {2011.10588},
 primaryClass = {astro-ph.SR},
       adsurl = {https://ui.adsabs.harvard.edu/abs/2021A&A...646A.119P}
}

@ARTICLE{Blagorodnova2021,
       author = {{Blagorodnova}, Nadejda and {Klencki}, Jakub and {Pejcha}, Ond{\v{r}}ej and {Vreeswijk}, Paul M. and {Bond}, Howard E. and {Burdge}, Kevin B. and {De}, Kishalay and {Fremling}, Christoffer and {Gehrz}, Robert D. and {Jencson}, Jacob E. and {Kasliwal}, Mansi M. and {Kupfer}, Thomas and {Lau}, Ryan M. and {Masci}, Frank J. and {Rich}, Michael R.},
        title = "{The luminous red nova AT 2018bwo in NGC 45 and its binary yellow supergiant progenitor}",
      journal = {\aap},
         year = 2021,
        month = sep,
       volume = {653},
          eid = {A134},
        pages = {A134},
          doi = {10.1051/0004-6361/202140525},
archivePrefix = {arXiv},
       eprint = {2102.05662},
 primaryClass = {astro-ph.SR},
       adsurl = {https://ui.adsabs.harvard.edu/abs/2021A&A...653A.134B}
}

@ARTICLE{Williams2015,
       author = {{Williams}, S.~C. and {Darnley}, M.~J. and {Bode}, M.~F. and {Steele}, I.~A.},
        title = "{A Luminous Red Nova in M31 and its Progenitor System}",
      journal = {\apjl},
         year = 2015,
        month = jun,
       volume = {805},
       number = {2},
          eid = {L18},
        pages = {L18},
          doi = {10.1088/2041-8205/805/2/L18},
archivePrefix = {arXiv},
       eprint = {1504.07747},
 primaryClass = {astro-ph.SR},
       adsurl = {https://ui.adsabs.harvard.edu/abs/2015ApJ...805L..18W}
}

@ARTICLE{Karambelkar2026,
       author = {{Karambelkar}, Viraj and {Kasliwal}, Mansi M. and {Lau}, Ryan M. and {Jencson}, Jacob E. and {Blagorodnova}, Nadejda and {G{\'o}mez-Mu{\~n}oz}, Marco A. and {Tranin}, Hugo and {Wavasseur}, Maxime and {Shahbandeh}, Melissa and {De}, Kishalay},
        title = "{Hot Springs and Dust Reservoirs: JWST Reveals the Dusty, Molecular Aftermath of Extragalactic Stellar Mergers}",
      journal = {\apj},
         year = 2026,
        month = mar,
       volume = {999},
       number = {1},
          eid = {16},
        pages = {16},
          doi = {10.3847/1538-4357/ae38bf},
archivePrefix = {arXiv},
       eprint = {2508.03932},
 primaryClass = {astro-ph.SR},
       adsurl = {https://ui.adsabs.harvard.edu/abs/2026ApJ...999...16K}
}

@ARTICLE{Reguitti2026,
       author = {{Reguitti}, A. and {Pastorello}, A. and {Valerin}, G.},
        title = "{The fate of the progenitors of luminous red novae: Infrared detection of LRNe years after the outburst}",
      journal = {\aap},
         year = 2026,
        month = feb,
       volume = {706},
          eid = {A154},
        pages = {A154},
          doi = {10.1051/0004-6361/202555226},
archivePrefix = {arXiv},
       eprint = {2504.14592},
 primaryClass = {astro-ph.SR},
       adsurl = {https://ui.adsabs.harvard.edu/abs/2026A&A...706A.154R}
}

@ARTICLE{2012MNRAS.427..127B,
       author = {{Bressan}, Alessandro and {Marigo}, Paola and {Girardi}, L{\'e}o. and {Salasnich}, Bernardo and {Dal Cero}, Claudia and {Rubele}, Stefano and {Nanni}, Ambra},
        title = "{PARSEC: stellar tracks and isochrones with the PAdova and TRieste Stellar Evolution Code}",
      journal = {\mnras},
         year = 2012,
        month = nov,
       volume = {427},
       number = {1},
        pages = {127-145},
          doi = {10.1111/j.1365-2966.2012.21948.x},
archivePrefix = {arXiv},
       eprint = {1208.4498},
 primaryClass = {astro-ph.SR},
       adsurl = {https://ui.adsabs.harvard.edu/abs/2012MNRAS.427..127B}
}

@ARTICLE{2014MNRAS.444.2525C,
       author = {{Chen}, Yang and {Girardi}, L{\'e}o and {Bressan}, Alessandro and {Marigo}, Paola and {Barbieri}, Mauro and {Kong}, Xu},
        title = "{Improving PARSEC models for very low mass stars}",
      journal = {\mnras},
         year = 2014,
        month = nov,
       volume = {444},
       number = {3},
        pages = {2525-2543},
          doi = {10.1093/mnras/stu1605},
archivePrefix = {arXiv},
       eprint = {1409.0322},
 primaryClass = {astro-ph.SR},
       adsurl = {https://ui.adsabs.harvard.edu/abs/2014MNRAS.444.2525C}
}

@ARTICLE{2015MNRAS.452.1068C,
       author = {{Chen}, Yang and {Bressan}, Alessandro and {Girardi}, L{\'e}o and {Marigo}, Paola and {Kong}, Xu and {Lanza}, Antonio},
        title = "{PARSEC evolutionary tracks of massive stars up to 350 M$_{{\ensuremath{\odot}}}$ at metallicities 0.0001 {\ensuremath{\leq}} Z {\ensuremath{\leq}} 0.04}",
      journal = {\mnras},
         year = 2015,
        month = sep,
       volume = {452},
       number = {1},
        pages = {1068-1080},
          doi = {10.1093/mnras/stv1281},
archivePrefix = {arXiv},
       eprint = {1506.01681},
 primaryClass = {astro-ph.SR},
       adsurl = {https://ui.adsabs.harvard.edu/abs/2015MNRAS.452.1068C}
}

@ARTICLE{2014MNRAS.445.4287T,
       author = {{Tang}, Jing and {Bressan}, Alessandro and {Rosenfield}, Philip and {Slemer}, Alessandra and {Marigo}, Paola and {Girardi}, L{\'e}o and {Bianchi}, Luciana},
        title = "{New PARSEC evolutionary tracks of massive stars at low metallicity: testing canonical stellar evolution in nearby star-forming dwarf galaxies}",
      journal = {\mnras},
         year = 2014,
        month = dec,
       volume = {445},
       number = {4},
        pages = {4287-4305},
          doi = {10.1093/mnras/stu2029},
archivePrefix = {arXiv},
       eprint = {1410.1745},
 primaryClass = {astro-ph.SR},
       adsurl = {https://ui.adsabs.harvard.edu/abs/2014MNRAS.445.4287T}
}

@ARTICLE{2017ApJ...835...77M,
       author = {{Marigo}, Paola and {Girardi}, L{\'e}o and {Bressan}, Alessandro and {Rosenfield}, Philip and {Aringer}, Bernhard and {Chen}, Yang and {Dussin}, Marco and {Nanni}, Ambra and {Pastorelli}, Giada and {Rodrigues}, Tha{\'\i}se S. and {Trabucchi}, Michele and {Bladh}, Sara and {Dalcanton}, Julianne and {Groenewegen}, Martin A.~T. and {Montalb{\'a}n}, Josefina and {Wood}, Peter R.},
        title = "{A New Generation of PARSEC-COLIBRI Stellar Isochrones Including the TP-AGB Phase}",
      journal = {\apj},
         year = 2017,
        month = jan,
       volume = {835},
       number = {1},
          eid = {77},
        pages = {77},
          doi = {10.3847/1538-4357/835/1/77},
archivePrefix = {arXiv},
       eprint = {1701.08510},
 primaryClass = {astro-ph.SR},
       adsurl = {https://ui.adsabs.harvard.edu/abs/2017ApJ...835...77M}
}

@ARTICLE{2019MNRAS.485.5666P,
       author = {{Pastorelli}, Giada and {Marigo}, Paola and {Girardi}, L{\'e}o and {Chen}, Yang and {Rubele}, Stefano and {Trabucchi}, Michele and {Aringer}, Bernhard and {Bladh}, Sara and {Bressan}, Alessandro and {Montalb{\'a}n}, Josefina and {Boyer}, Martha L. and {Dalcanton}, Julianne J. and {Eriksson}, Kjell and {Groenewegen}, Martin A.~T. and {H{\"o}fner}, Susanne and {Lebzelter}, Thomas and {Nanni}, Ambra and {Rosenfield}, Philip and {Wood}, Peter R. and {Cioni}, Maria-Rosa L.},
        title = "{Constraining the thermally pulsing asymptotic giant branch phase with resolved stellar populations in the Small Magellanic Cloud}",
      journal = {\mnras},
         year = 2019,
        month = jun,
       volume = {485},
       number = {4},
        pages = {5666-5692},
          doi = {10.1093/mnras/stz725},
archivePrefix = {arXiv},
       eprint = {1903.04499},
 primaryClass = {astro-ph.SR},
       adsurl = {https://ui.adsabs.harvard.edu/abs/2019MNRAS.485.5666P}
}

@ARTICLE{2020MNRAS.498.3283P,
       author = {{Pastorelli}, Giada and {Marigo}, Paola and {Girardi}, L{\'e}o and {Aringer}, Bernhard and {Chen}, Yang and {Rubele}, Stefano and {Trabucchi}, Michele and {Bladh}, Sara and {Boyer}, Martha L. and {Bressan}, Alessandro and {Dalcanton}, Julianne J. and {Groenewegen}, Martin A.~T. and {Lebzelter}, Thomas and {Mowlavi}, Nami and {Chubb}, Katy L. and {Cioni}, Maria-Rosa L. and {de Grijs}, Richard and {Ivanov}, Valentin D. and {Nanni}, Ambra and {van Loon}, Jacco Th and {Zaggia}, Simone},
        title = "{Constraining the thermally pulsing asymptotic giant branch phase with resolved stellar populations in the Large Magellanic Cloud}",
      journal = {\mnras},
         year = 2020,
        month = nov,
       volume = {498},
       number = {3},
        pages = {3283-3301},
          doi = {10.1093/mnras/staa2565},
archivePrefix = {arXiv},
       eprint = {2008.08595},
 primaryClass = {astro-ph.SR},
       adsurl = {https://ui.adsabs.harvard.edu/abs/2020MNRAS.498.3283P}
}

@ARTICLE{De2026,
       author = {{De}, Kishalay and {MacLeod}, Morgan and {Jencson}, Jacob E. and {Lau}, Ryan M. and {Antoni}, Andrea and {Colmenares}, Mar{\'\i}a Jos{\'e} and {Huang}, Jane and {Masterson}, Megan and {Karambelkar}, Viraj R. and {Kasliwal}, Mansi M. and {Loeb}, Abraham and {Panagiotou}, Christos and {Quataert}, Eliot},
        title = "{Fading into Darkness: A Weak Mass Ejection and Low-efficiency Fallback Accompanying Black Hole Formation in M31-2014-DS1}",
      journal = {\apjl},
         year = 2026,
        month = mar,
       volume = {999},
       number = {2},
          eid = {L25},
        pages = {L25},
          doi = {10.3847/2041-8213/ae468d},
archivePrefix = {arXiv},
       eprint = {2601.05774},
 primaryClass = {astro-ph.SR},
       adsurl = {https://ui.adsabs.harvard.edu/abs/2026ApJ...999L..25D}
}

@ARTICLE{Kochanek2024dusty,
       author = {{Kochanek}, C.~S.},
        title = "{Transients obscured by dusty discs}",
      journal = {\mnras},
         year = 2024,
        month = apr,
       volume = {529},
       number = {3},
        pages = {1958-1969},
          doi = {10.1093/mnras/stae589},
archivePrefix = {arXiv},
       eprint = {2305.11936},
 primaryClass = {astro-ph.SR},
       adsurl = {https://ui.adsabs.harvard.edu/abs/2024MNRAS.529.1958K}
}

@ARTICLE{De2026Science,
       author = {{De}, Kishalay and {MacLeod}, Morgan and {Jencson}, Jacob E. and {Lovegrove}, Elizabeth and {Antoni}, Andrea and {Kara}, Erin and {Kasliwal}, Mansi M. and {Lau}, Ryan M. and {Loeb}, Abraham and {Masterson}, Megan and {Meisner}, Aaron M. and {Panagiotou}, Christos and {Quataert}, Eliot and {Simcoe}, Robert},
        title = "{Disappearance of a massive star in the Andromeda Galaxy due to formation of a black hole}",
      journal = {Science},
         year = 2026,
        month = feb,
       volume = {391},
       number = {6786},
        pages = {689-693},
          doi = {10.1126/science.adt4853},
archivePrefix = {arXiv},
       eprint = {2410.14778},
 primaryClass = {astro-ph.HE},
       adsurl = {https://ui.adsabs.harvard.edu/abs/2026Sci...391..689D}
}

@ARTICLE{Gerke2015,
       author = {{Gerke}, J.~R. and {Kochanek}, C.~S. and {Stanek}, K.~Z.},
        title = "{The search for failed supernovae with the Large Binocular Telescope: first candidates}",
      journal = {\mnras},
         year = 2015,
        month = jul,
       volume = {450},
       number = {3},
        pages = {3289-3305},
          doi = {10.1093/mnras/stv776},
archivePrefix = {arXiv},
       eprint = {1411.1761},
 primaryClass = {astro-ph.SR},
       adsurl = {https://ui.adsabs.harvard.edu/abs/2015MNRAS.450.3289G}
}

@ARTICLE{VanDyk2012,
       author = {{Van Dyk}, Schuyler D. and {Davidge}, Tim J. and {Elias-Rosa}, Nancy and {Taubenberger}, Stefan and {Li}, Weidong and {Levesque}, Emily M. and {Howerton}, Stanley and {Pignata}, Giuliano and {Morrell}, Nidia and {Hamuy}, Mario and {Filippenko}, Alexei V.},
        title = "{Supernova 2008bk and Its Red Supergiant Progenitor}",
      journal = {\aj},
         year = 2012,
        month = jan,
       volume = {143},
       number = {1},
          eid = {19},
        pages = {19},
          doi = {10.1088/0004-6256/143/1/19},
archivePrefix = {arXiv},
       eprint = {1011.5873},
 primaryClass = {astro-ph.SR},
       adsurl = {https://ui.adsabs.harvard.edu/abs/2012AJ....143...19V}
}

@ARTICLE{Maund2014,
       author = {{Maund}, Justyn R. and {Reilly}, Emma and {Mattila}, Seppo},
        title = "{A late-time view of the progenitors of five Type IIP supernovae}",
      journal = {\mnras},
         year = 2014,
        month = feb,
       volume = {438},
       number = {2},
        pages = {938-958},
          doi = {10.1093/mnras/stt2131},
archivePrefix = {arXiv},
       eprint = {1302.7152},
 primaryClass = {astro-ph.SR},
       adsurl = {https://ui.adsabs.harvard.edu/abs/2014MNRAS.438..938M}
}

@ARTICLE{Burrows2021,
       author = {{Burrows}, A. and {Vartanyan}, D.},
        title = "{Core-collapse supernova explosion theory}",
      journal = {\nat},
         year = 2021,
        month = jan,
       volume = {589},
       number = {7840},
        pages = {29-39},
          doi = {10.1038/s41586-020-03059-w},
archivePrefix = {arXiv},
       eprint = {2009.14157},
 primaryClass = {astro-ph.SR},
       adsurl = {https://ui.adsabs.harvard.edu/abs/2021Natur.589...29B}
}

@ARTICLE{Mezzacappa2005,
       author = {{Mezzacappa}, Anthony},
        title = "{ASCERTAINING THE CORE COLLAPSE SUPERNOVA MECHANISM: The State of the Art and the Road Ahead}",
      journal = {Annual Review of Nuclear and Particle Science},
         year = 2005,
        month = dec,
       volume = {55},
       number = {1},
        pages = {467-515},
          doi = {10.1146/annurev.nucl.55.090704.151608},
       adsurl = {https://ui.adsabs.harvard.edu/abs/2005ARNPS..55..467M}
}

@ARTICLE{Smartt2015,
       author = {{Smartt}, S.~J.},
        title = "{Observational Constraints on the Progenitors of Core-Collapse Supernovae: The Case for Missing High-Mass Stars}",
      journal = {\pasa},
         year = 2015,
        month = apr,
       volume = {32},
          eid = {e016},
        pages = {e016},
          doi = {10.1017/pasa.2015.17},
archivePrefix = {arXiv},
       eprint = {1504.02635},
 primaryClass = {astro-ph.SR},
       adsurl = {https://ui.adsabs.harvard.edu/abs/2015PASA...32...16S}
}

@ARTICLE{Botticella2012,
       author = {{Botticella}, M.~T. and {Smartt}, S.~J. and {Kennicutt}, R.~C. and {Cappellaro}, E. and {Sereno}, M. and {Lee}, J.~C.},
        title = "{A comparison between star formation rate diagnostics and rate of core collapse supernovae within 11 Mpc}",
      journal = {\aap},
         year = 2012,
        month = jan,
       volume = {537},
          eid = {A132},
        pages = {A132},
          doi = {10.1051/0004-6361/201117343},
archivePrefix = {arXiv},
       eprint = {1111.1692},
 primaryClass = {astro-ph.CO},
       adsurl = {https://ui.adsabs.harvard.edu/abs/2012A&A...537A.132B}
}

@ARTICLE{LVK2025,
       author = {{The LIGO Scientific Collaboration} and {the Virgo Collaboration} and {the KAGRA Collaboration}},
        title = "{GWTC-4.0: Updating the Gravitational-Wave Transient Catalog with Observations from the First Part of the Fourth LIGO-Virgo-KAGRA Observing Run}",
      journal = {arXiv e-prints},
         year = 2025,
        month = aug,
          eid = {arXiv:2508.18082},
        pages = {arXiv:2508.18082},
          doi = {10.48550/arXiv.2508.18082},
archivePrefix = {arXiv},
       eprint = {2508.18082},
 primaryClass = {gr-qc},
       adsurl = {https://ui.adsabs.harvard.edu/abs/2025arXiv250818082T}
}

@ARTICLE{Ugolini2025,
       author = {{Ugolini}, Cristiano and {Limongi}, Marco and {Schneider}, Raffaella and {Chieffi}, Alessandro and {Di Carlo}, Ugo Niccol{\`o} and {Spera}, Mario},
        title = "{The initial mass-remnant mass relation for core collapse supernovae}",
      journal = {\aap},
         year = 2025,
        month = mar,
       volume = {695},
          eid = {A122},
        pages = {A122},
          doi = {10.1051/0004-6361/202451483},
archivePrefix = {arXiv},
       eprint = {2501.18689},
 primaryClass = {astro-ph.HE},
       adsurl = {https://ui.adsabs.harvard.edu/abs/2025A&A...695A.122U}
}

@ARTICLE{OConnor2011,
       author = {{O'Connor}, Evan and {Ott}, Christian D.},
        title = "{Black Hole Formation in Failing Core-Collapse Supernovae}",
      journal = {\apj},
         year = 2011,
        month = apr,
       volume = {730},
       number = {2},
          eid = {70},
        pages = {70},
          doi = {10.1088/0004-637X/730/2/70},
archivePrefix = {arXiv},
       eprint = {1010.5550},
 primaryClass = {astro-ph.HE},
       adsurl = {https://ui.adsabs.harvard.edu/abs/2011ApJ...730...70O}
}

@ARTICLE{Ertl2016,
       author = {{Ertl}, T. and {Janka}, H.-Th. and {Woosley}, S.~E. and {Sukhbold}, T. and {Ugliano}, M.},
        title = "{A Two-parameter Criterion for Classifying the Explodability of Massive Stars by the Neutrino-driven Mechanism}",
      journal = {\apj},
         year = 2016,
        month = feb,
       volume = {818},
       number = {2},
          eid = {124},
        pages = {124},
          doi = {10.3847/0004-637X/818/2/124},
archivePrefix = {arXiv},
       eprint = {1503.07522},
 primaryClass = {astro-ph.SR},
       adsurl = {https://ui.adsabs.harvard.edu/abs/2016ApJ...818..124E}
}

@ARTICLE{Sukhbold2016,
       author = {{Sukhbold}, Tuguldur and {Ertl}, T. and {Woosley}, S.~E. and {Brown}, Justin M. and {Janka}, H.-T.},
        title = "{Core-collapse Supernovae from 9 to 120 Solar Masses Based on Neutrino-powered Explosions}",
      journal = {\apj},
         year = 2016,
        month = apr,
       volume = {821},
       number = {1},
          eid = {38},
        pages = {38},
          doi = {10.3847/0004-637X/821/1/38},
archivePrefix = {arXiv},
       eprint = {1510.04643},
 primaryClass = {astro-ph.HE},
       adsurl = {https://ui.adsabs.harvard.edu/abs/2016ApJ...821...38S}
}

@ARTICLE{Ugliano2012,
       author = {{Ugliano}, Marcella and {Janka}, Hans-Thomas and {Marek}, Andreas and {Arcones}, Almudena},
        title = "{Progenitor-explosion Connection and Remnant Birth Masses for Neutrino-driven Supernovae of Iron-core Progenitors}",
      journal = {\apj},
         year = 2012,
        month = sep,
       volume = {757},
       number = {1},
          eid = {69},
        pages = {69},
          doi = {10.1088/0004-637X/757/1/69},
archivePrefix = {arXiv},
       eprint = {1205.3657},
 primaryClass = {astro-ph.SR},
       adsurl = {https://ui.adsabs.harvard.edu/abs/2012ApJ...757...69U}
}

@ARTICLE{Luo2025,
       author = {{Luo}, Renyu and {Zhu}, Chunhua and {L{\"u}}, Guoliang and {Liu}, Helei and {Guo}, Sufen and {Li}, Lei and {Li}, Zhuowen},
        title = "{Effect of rotation and metallicity on the explodability of massive stars}",
      journal = {\aap},
         year = 2025,
        month = nov,
       volume = {704},
          eid = {A46},
        pages = {A46},
          doi = {10.1051/0004-6361/202555081},
archivePrefix = {arXiv},
       eprint = {2510.06043},
 primaryClass = {astro-ph.SR},
       adsurl = {https://ui.adsabs.harvard.edu/abs/2025A&A...704A..46L}
}

@ARTICLE{Antoni2023,
    author = {Antoni, Andrea and Quataert, Eliot},
    title = {Numerical simulations of the random angular momentum in convection – II. Delayed explosions of red supergiants following ‘failed’ supernovae},
    journal = {\mnras},
    volume = {525},
    number = {1},
    pages = {1229-1245},
    year = {2023},
    month = {08},
    issn = {0035-8711},
    doi = {10.1093/mnras/stad2328},
    url = {https://doi.org/10.1093/mnras/stad2328},
    eprint = {https://academic.oup.com/mnras/article-pdf/525/1/1229/51123464/stad2328.pdf},
}

@ARTICLE{Ivanov2021,
       author = {{Ivanov}, Mario and {Fern{\'a}ndez}, Rodrigo},
        title = "{Mass Ejection in Failed Supernovae: Equation of State and Neutrino Loss Dependence}",
      journal = {\apj},
         year = 2021,
        month = apr,
       volume = {911},
       number = {1},
          eid = {6},
        pages = {6},
          doi = {10.3847/1538-4357/abe59e},
archivePrefix = {arXiv},
       eprint = {2101.02712},
 primaryClass = {astro-ph.HE},
       adsurl = {https://ui.adsabs.harvard.edu/abs/2021ApJ...911....6I}
}

@ARTICLE{Fernandez2018,
       author = {{Fern{\'a}ndez}, Rodrigo and {Quataert}, Eliot and {Kashiyama}, Kazumi and {Coughlin}, Eric R.},
        title = "{Mass ejection in failed supernovae: variation with stellar progenitor}",
      journal = {\mnras},
         year = 2018,
        month = may,
       volume = {476},
       number = {2},
        pages = {2366-2383},
          doi = {10.1093/mnras/sty306},
archivePrefix = {arXiv},
       eprint = {1710.01735},
 primaryClass = {astro-ph.HE},
       adsurl = {https://ui.adsabs.harvard.edu/abs/2018MNRAS.476.2366F}
}

@ARTICLE{Lovegrove2017,
       author = {{Lovegrove}, Elizabeth and {Woosley}, S.~E. and {Zhang}, Weiqun},
        title = "{Very Low-energy Supernovae: Light Curves and Spectra of Shock Breakout}",
      journal = {\apj},
         year = 2017,
        month = aug,
       volume = {845},
       number = {2},
          eid = {103},
        pages = {103},
          doi = {10.3847/1538-4357/aa7b7d},
archivePrefix = {arXiv},
       eprint = {1706.02440},
 primaryClass = {astro-ph.HE},
       adsurl = {https://ui.adsabs.harvard.edu/abs/2017ApJ...845..103L}
}

@ARTICLE{Lovegrove2013,
       author = {{Lovegrove}, Elizabeth and {Woosley}, S.~E.},
        title = "{Very Low Energy Supernovae from Neutrino Mass Loss}",
      journal = {\apj},
         year = 2013,
        month = jun,
       volume = {769},
       number = {2},
          eid = {109},
        pages = {109},
          doi = {10.1088/0004-637X/769/2/109},
archivePrefix = {arXiv},
       eprint = {1303.5055},
 primaryClass = {astro-ph.HE},
       adsurl = {https://ui.adsabs.harvard.edu/abs/2013ApJ...769..109L}
}

@ARTICLE{Vartanyan2023,
       author = {{Vartanyan}, David and {Burrows}, Adam and {Wang}, Tianshu and {Coleman}, Matthew S.~B. and {White}, Christopher J.},
        title = "{Gravitational-wave signature of core-collapse supernovae}",
      journal = {\prd},
         year = 2023,
        month = may,
       volume = {107},
       number = {10},
          eid = {103015},
        pages = {103015},
          doi = {10.1103/PhysRevD.107.103015},
archivePrefix = {arXiv},
       eprint = {2302.07092},
 primaryClass = {astro-ph.HE},
       adsurl = {https://ui.adsabs.harvard.edu/abs/2023PhRvD.107j3015V}
}

@ARTICLE{Powell2025,
       author = {{Powell}, Jade and {M{\"u}ller}, Bernhard},
        title = "{Gravitational waves from core-collapse supernovae with no electromagnetic counterparts}",
      journal = {Classical and Quantum Gravity},
         year = 2025,
        month = nov,
       volume = {42},
       number = {21},
          eid = {215002},
        pages = {215002},
          doi = {10.1088/1361-6382/ae0dea},
archivePrefix = {arXiv},
       eprint = {2506.03581},
 primaryClass = {astro-ph.HE},
       adsurl = {https://ui.adsabs.harvard.edu/abs/2025CQGra..42u5002P}
}

@ARTICLE{Liebendorfer2004,
       author = {{Liebend{\"o}rfer}, Matthias and {Messer}, O.~E. Bronson and {Mezzacappa}, Anthony and {Bruenn}, Stephen W. and {Cardall}, Christian Y. and {Thielemann}, F.-K.},
        title = "{A Finite Difference Representation of Neutrino Radiation Hydrodynamics in Spherically Symmetric General Relativistic Spacetime}",
      journal = {\apjs},
         year = 2004,
        month = jan,
       volume = {150},
       number = {1},
        pages = {263-316},
          doi = {10.1086/380191},
archivePrefix = {arXiv},
       eprint = {astro-ph/0207036},
 primaryClass = {astro-ph},
       adsurl = {https://ui.adsabs.harvard.edu/abs/2004ApJS..150..263L}
}

@ARTICLE{Kuroda2023,
    author = {Kuroda, Takami and Shibata, Masaru},
    title = {Failed supernova simulations beyond black hole formation},
    journal = {\mnras},
    volume = {526},
    number = {1},
    pages = {152-159},
    year = {2023},
    month = {09},
    issn = {0035-8711},
    doi = {10.1093/mnras/stad2710},
    url = {https://doi.org/10.1093/mnras/stad2710},
    eprint = {https://academic.oup.com/mnras/article-pdf/526/1/152/51722610/stad2710.pdf},
}

@ARTICLE{Nakanishi2026,
       author = {{Nakanishi}, F. and {Abe}, K. and {Abe}, S. and {Asaoka}, Y. and {Harada}, M. and {Hayato}, Y. and {Hiraide}, K. and {Hosokawa}, K. and {Hung}, T.~H. and {Ieki}, K. and {Ikeda}, M. and {Kameda}, J. and {Kanemura}, Y. and {Kataoka}, Y. and {Miki}, S. and {Mine}, S.},
        title = "{First Associated Neutrino Search for a Failed Supernova Candidate with Super-Kamiokande}",
      journal = {\apjl},
         year = 2026,
        month = jan,
       volume = {997},
       number = {1},
          eid = {L9},
        pages = {L9},
          doi = {10.3847/2041-8213/ae2c73},
archivePrefix = {arXiv},
       eprint = {2511.03470},
 primaryClass = {astro-ph.HE},
       adsurl = {https://ui.adsabs.harvard.edu/abs/2026ApJ...997L...9N}
}

@ARTICLE{Kochanek2008,
       author = {{Kochanek}, Christopher S. and {Beacom}, John F. and {Kistler}, Matthew D. and {Prieto}, Jos{\'e} L. and {Stanek}, Krzysztof Z. and {Thompson}, Todd A. and {Y{\"u}ksel}, Hasan},
        title = "{A Survey About Nothing: Monitoring a Million Supergiants for Failed Supernovae}",
      journal = {\apj},
         year = 2008,
        month = sep,
       volume = {684},
       number = {2},
        pages = {1336-1342},
          doi = {10.1086/590053},
archivePrefix = {arXiv},
       eprint = {0802.0456},
 primaryClass = {astro-ph},
       adsurl = {https://ui.adsabs.harvard.edu/abs/2008ApJ...684.1336K}
}

@ARTICLE{Neustadt2021,
    author = {Neustadt, J M M and Kochanek, C S and Stanek, K Z and Basinger, C and Jayasinghe, T and Garling, C T and Adams, S M and Gerke, J},
    title = {The search for failed supernovae with the Large Binocular Telescope: a new candidate and the failed SN fraction with 11 yr of data},
    journal = {\mnras},
    volume = {508},
    number = {1},
    pages = {516-528},
    year = {2021},
    month = {09},
    issn = {0035-8711},
    doi = {10.1093/mnras/stab2605},
    url = {https://doi.org/10.1093/mnras/stab2605},
    eprint = {https://academic.oup.com/mnras/article-pdf/508/1/516/40440715/stab2605.pdf},
}

@ARTICLE{Adams2017search,
    author = {Adams, S. M. and Kochanek, C. S. and Gerke, J. R. and Stanek, K. Z.},
    title = {The search for failed supernovae with the Large Binocular Telescope: constraints from 7 yr of data},
    journal = {\mnras},
    volume = {469},
    number = {2},
    pages = {1445-1455},
    year = {2017a},
    month = {04},
    issn = {0035-8711},
    doi = {10.1093/mnras/stx898},
    url = {https://doi.org/10.1093/mnras/stx898},
    eprint = {https://academic.oup.com/mnras/article-pdf/469/2/1445/17272527/stx898.pdf},
}

@ARTICLE{Kochanek2014,
       author = {{Kochanek}, C.~S. and {Adams}, Scott M. and {Belczynski}, Krzysztof},
        title = "{Stellar mergers are common}",
      journal = {\mnras},
         year = 2014,
        month = sep,
       volume = {443},
       number = {2},
        pages = {1319-1328},
          doi = {10.1093/mnras/stu1226},
archivePrefix = {arXiv},
       eprint = {1405.1042},
 primaryClass = {astro-ph.SR},
       adsurl = {https://ui.adsabs.harvard.edu/abs/2014MNRAS.443.1319K}
}

@ARTICLE{Davies2020-1,
       author = {{Davies}, Ben and {Beasor}, Emma R.},
        title = "{The `red supergiant problem': the upper luminosity boundary of Type II supernova progenitors}",
      journal = {\mnras},
         year = 2020,
        month = mar,
       volume = {493},
       number = {1},
        pages = {468-476},
          doi = {10.1093/mnras/staa174},
archivePrefix = {arXiv},
       eprint = {2001.06020},
 primaryClass = {astro-ph.SR},
       adsurl = {https://ui.adsabs.harvard.edu/abs/2020MNRAS.493..468D}
}

@ARTICLE{Kochanek2020,
       author = {{Kochanek}, C.~S.},
        title = "{On the red supergiant problem}",
      journal = {\mnras},
         year = 2020,
        month = apr,
       volume = {493},
       number = {4},
        pages = {4945-4949},
          doi = {10.1093/mnras/staa605},
archivePrefix = {arXiv},
       eprint = {2001.07216},
 primaryClass = {astro-ph.SR},
       adsurl = {https://ui.adsabs.harvard.edu/abs/2020MNRAS.493.4945K}
}

\end{document}